\documentclass[12pt]{spieman}

\usepackage{amsmath,amsfonts,amssymb}
\usepackage{graphicx}
\usepackage[table]{xcolor}
\usepackage{array}
\usepackage{booktabs}
\usepackage{tabularx}
\usepackage{longtable}
\usepackage{makecell}
\usepackage{enumitem}
\usepackage{lscape}              
\usepackage{textcomp}
\usepackage{tocloft}
\usepackage{lineno}                
\usepackage{lastpage}

\definecolor{esablue}{HTML}{1F4E79}
\definecolor{rowalt}{HTML}{EAF1F8}
\definecolor{section}{HTML}{D5E3EF}
\definecolor{leo}{HTML}{FFF6E0}
\definecolor{hwo}{HTML}{E6F0E6}
\definecolor{trlfull}{HTML}{C8E6C9}
\definecolor{trlpart}{HTML}{FCEABB}
\definecolor{trlnone}{HTML}{F2F2F2}
\definecolor{leohdr}{HTML}{B8860B}
\definecolor{hwohdr}{HTML}{2E7D32}
\definecolor{uvbox}{HTML}{F0E5F5}
\definecolor{visbox}{HTML}{E5F0F5}
\definecolor{nirbox}{HTML}{E8F5E5}

\renewcommand{\arraystretch}{1.25}
\newcolumntype{P}[1]{>{\raggedright\arraybackslash}p{#1}}
\newcolumntype{C}[1]{>{\centering\arraybackslash}p{#1}}

\newcommand{\cellbullets}[1]{%
  \begingroup
  \setlist[itemize]{nosep,leftmargin=*,topsep=0pt,parsep=0pt,partopsep=0pt,label={\textbullet}}%
  \footnotesize
  \begin{itemize}#1\end{itemize}%
  \endgroup
}

\newcommand{\trlfull}{\cellcolor{trlfull}\textbf{$\bullet$}}
\newcommand{\trlpart}{\cellcolor{trlpart}\textbf{$\circ$}}
\newcommand{\trlhalf}{\cellcolor{trlpart}\textbf{$\circledcirc$}}
\newcommand{\trlnone}{\cellcolor{trlnone}\textbf{--}}

\cftpagenumbersoff{figure}
\cftpagenumbersoff{table}

\title{Qualification Pathways for Photonic Integrated Circuits in Astrophotonic Space Missions}

\author[a,*]{Kalaga Madhav}
\author[a]{Aashia Rahman}
\author[b]{Máté Ádámkovics}
\affil[a]{Astrophotonics (innoFSPEC), Leibniz-Institut f\"ur Astrophysik Potsdam (AIP),
          An der Sternwarte 16, 14482 Potsdam, Germany}
\affil[b]{Lockheed Martin Advanced Technology Center, Palo Alto, California, USA}

\begin{document}
\maketitle

\begin{abstract}
Photonic integrated circuits (PICs) promise order-of-magnitude reductions in the size, weight and power (SWaP) of optical subsystems for astronomy, planetary and Earth-observation missions, yet no PIC-specific space-qualification standard exists. This paper consolidates the principal NASA and ESA qualification documents that apply, or can be tailored, to astrophotonic PICs --- arrayed waveguide gratings, photonic lanterns, fibre Bragg gratings, and integrated beam combiners (ABCD, discrete beam combiners, nullers) for spectrographs and stellar interferometers. A master qualification table lists 19 standard test steps with applicable standards and EU/USA test facilities. Two reference mission profiles --- a LEO smallsat demonstrator and an HWO-class Lagrange-2 flagship --- yield a tailoring matrix, while a TRL-versus-test-coverage roadmap maps each activity onto the NASA/ESA readiness levels and review gates. A survey of UV/visible/near-infrared platforms relates spectral coverage, maturity and flight heritage, and a radiation-effects summary shows passive silica, Si\textsubscript{3}N\textsubscript{4} and laser-written cores are essentially radiation-tolerant while active III--V and Ge devices carry the hardness burden. The central outcome is a seven-phase qualification template (PIC-SQT) with explicit TRL gates, exact test procedures and mission-class tailoring; we further identify qualification processes relevant to PICs that current standards do not cover, and document 40+ years of optical-fibre flight heritage.
\end{abstract}

\keywords{photonic integrated circuits, astrophotonics, space qualification,
radiation hardness, NASA standards, ESA ECSS standards}

{\noindent \footnotesize\textbf{*}Kalaga Madhav,
 \linkable{kmadhav@aip.de}}

\begin{spacing}{1}

\tableofcontents
\vspace{1em}

\section{Introduction}\label{sec:intro}
Modern space missions in astronomy, planetary science and Earth observation
are simultaneously photon-starved and data-rich: they demand ever larger
collecting apertures and ever higher spectral, spatial and temporal
information content, yet are bounded by the mass, volume and power (SWaP)
that a launch vehicle and a platform can carry. Photonic integrated circuits
(PICs) --- which implement on a millimetre-scale chip the spectral filtering,
beam combination, routing, modulation and detection functions that
conventionally occupy a benchtop of bulk optics or a rack of fibre
components --- offer a direct route past this bottleneck. By replacing
discrete, individually aligned optical elements with lithographically
defined, wafer-scale, monolithically aligned circuits, PICs can collapse the
SWaP budget of an optical subsystem by one to two orders of magnitude while
removing the thermo-mechanical alignment drift that dominates the error
budget of bulk instruments~\cite{jovanovic2023,terrasanta2025}.

This advantage is not merely convenient; for space telescopes it is
structural. In a conventional instrument the size of the optical train ---
collimator, disperser and camera --- scales with the telescope diameter $D$,
so that instrument volume, mass and cost grow roughly as
$D^{3}$~\cite{blandhawthorn2009}. On the ground this cubic scaling is
affordable: the ELT first-light spectrograph HARMONI, for example, stands
8\,m tall and weighs some 40\,tonnes, yet sits comfortably on a Nasmyth
platform, and the complexity and cost it represents are managed with civil
infrastructure. In space the same scaling collides with a hard wall. Launch
mass and fairing volume are fixed by the rocket --- JWST flew at a
$\sim$6.2\,tonne launch mass folded into a 4.5\,m-class fairing --- and they
do \emph{not} grow with aperture. A 6\,m-class flagship such as the Habitable
Worlds Observatory therefore cannot carry a $D^{3}$-scaled conventional
instrument suite: increasing the primary-mirror diameter forces the
instrument SWaP budget to be \emph{reallocated} --- traded against the mirror,
sunshield, structure and propellant --- rather than allowed to expand with
the telescope. Photonic integrated circuits break this scaling at its root. A
single-mode photonic spectrograph or beam combiner has a size set by the
operating wavelength rather than by the telescope diameter, so its SWaP is
essentially decoupled from aperture~\cite{blandhawthorn2009,jovanovic2023};
the seeing- or diffraction-limited telescope feed is reduced to single mode by
a photonic lantern, after which all downstream processing happens on a chip
whose dimensions no longer track $D$. This decoupling is what makes PICs not
merely attractive but \emph{enabling} for the next generation of large space
telescopes.

The SWaP argument is concrete and already demonstrated across the three
mission domains this paper addresses. \emph{In astronomy}, the field of
astrophotonics was founded on exactly this premise~\cite{blandhawthorn2009}:
an arrayed-waveguide-grating (AWG) spectrograph disperses light on a
centimetre-scale chip rather than a metre-scale bench~\cite{gatkine2017,gatkine2022};
a photonic lantern converts a seeing-limited multimode telescope feed into an
array of diffraction-limited single-mode channels~\cite{leonsaval2010}; and
integrated beam combiners and nullers perform on a chip the interferometric
recombination that otherwise needs a vibration-isolated optical
table~\cite{norris2020glint,martinod2021glint}. Instrument-level
demonstrations such as the Potsdam Arrayed Waveguide Spectrograph and its
companion microcomb calibrator show that an entire near-infrared spectrograph
can be reduced to a cooled chip with a fibre feed~\cite{madhav2024,roth2023spie}
--- a decisive advantage for the photon-starved, stability-critical
instruments of flagships such as the Habitable Worlds Observatory (HWO),
whose $10^{-10}$ exoplanet-imaging contrast budget rewards the monolithic
phase stability of a PIC nuller~\cite{jovanovic2023,nasaHWO}.

\emph{In planetary and deep-space missions}, optical communication is
displacing radio for high-rate links precisely because of SWaP: a chip- and
fibre-based optical terminal delivers orders-of-magnitude higher data rate
per kilogram and per watt than an equivalent radio-frequency system. NASA's
Deep Space Optical Communications terminal on \emph{Psyche} returned
high-rate data from beyond 1\,AU~\cite{biswas2024dsoc,nasaDSOC}, while the 6U
TBIRD CubeSat downlinked 200\,Gbps using commercial PIC-derived coherent
transceivers~\cite{schieler2023} --- a data rate per unit mass unattainable
with RF, and one built from the same integrated laser, modulator and detector
blocks that feed planetary laser altimeters and spectrometers.
\emph{In Earth observation}, coherent frequency-modulated continuous-wave
(FMCW) LiDAR for ranging, wind sensing and proximity operations is converging
on integrated photonic transmitter/receiver engines~\cite{martin2018,lihachev2024},
and inter-satellite and space-to-ground optical links increasingly rely on
integrated InP, silicon and silicon-nitride transceivers whose SWaP and
radiation behaviour have been reviewed for the satellite-communication
community~\cite{terrasanta2025}. Across all three domains the trajectory is
identical: functions that were once bulk optics or discrete fibre assemblies
are migrating onto chips, with compounding gains in mass, volume, power,
alignment stability and --- through wafer-scale replication --- unit cost.

This very advantage, however, creates a problem. Astrophotonic PICs are novel
devices that do not yet belong to any closed family of space-qualified
electrical, electronic and electromechanical (EEE) or optoelectronic parts,
and no PIC-specific space-qualification standard exists. The qualification
documentation that programmes must use --- the ESA ECSS series, NASA
GSFC-STD-7000B GEVS and EEE-INST-002, and the Telcordia, MIL and JEDEC
industry methods --- was written for discrete components, hermetic packages
and optical fibres; it addresses neither the integrated nature of a PIC nor
the photon-starved, phase-stability-critical regime of astronomical
instruments. Both agencies explicitly permit \emph{tailoring} of their flows
for novel devices, but tailoring presupposes a coherent framework that does
not yet exist. In its absence each project re-derives an ad-hoc flow, and the
resulting fragmentation is itself a barrier to flying PICs and to the
SWaP-driven science they enable~\cite{jovanovic2023}.

This paper develops that missing framework. It compiles the applicable ESA
and NASA standards into a single master qualification flow, bounds the flow
between two reference mission classes through a tailoring matrix, maps each
activity onto the agencies' technology-readiness levels and review gates,
surveys the candidate material platforms and their radiation behaviour, and
consolidates the four decades of optical-fibre flight heritage that defines
the PIC-to-instrument interface. From these elements it distils a reusable,
seven-phase space-qualification template (PIC-SQT) with explicit test
procedures and mission-class tailoring, and it identifies the qualification
processes that remain uncovered by current standards. The remainder of the
paper is organised accordingly, beginning with its scope and purpose below.

\section{Scope and purpose}
This document compiles the principal qualification standards issued by NASA and the European Space Agency (ESA) that apply, or can be tailored, to Photonic Integrated Circuits (PICs) intended for space-based astrophotonic instruments — in particular arrayed waveguide gratings (AWGs), photonic lanterns, fibre Bragg gratings, and integrated beam combiners (ABCD, discrete beam combiners, nullers) for spectrographs and stellar interferometers.

Astrophotonic PICs do not currently belong to a closed family of EEE qualified parts. Both ESA (through ECSS-Q-ST-60C and the ESA/SCC system) and NASA (through GSFC-STD-7000 GEVS and EEE-INST-002, soon NASA-STD-8739.11) explicitly allow tailoring of EEE/ optoelectronic qualification flows for novel devices. This paper attempts to create a framework of differences clearly, rather than assuming that one qualification route applies equally to all technologies, especially in the context of photon-starved applications in astronomy. The qualification flow shown here therefore combines:

\begin{itemize}[leftmargin=1.4em]
  \item \textbf{ESA standards}: ECSS-Q-ST-60C Rev.4, ECSS-Q-ST-60-13C Rev.2 (Commercial EEE), ECSS-Q-ST-60-15C Rev.1 (RHA, Mar 2025), ECSS-Q-ST-70-02C, ECSS-Q-ST-70-04C, ECSS-Q-ST-70-06C, and the ESA/SCC Generic and Detail Specifications system maintained by the ESA Photonic Components Section.
  \item \textbf{NASA standards}: GSFC-STD-7000B GEVS (April 2021), EEE-INST-002 with Addendum 1, the forthcoming NASA-STD-8739.11 EEEE Parts Selection, Testing and Derating Standard (adds Parts Assurance Level 4 commercial option), NPR 8735.1 and NEPP guidance.
  \item \textbf{Industry/military test methods}: MIL-STD-883, MIL-STD-750, JEDEC JS-001/JS-002/JESD57, ASTM E595/E512, IEC 60068, and Telcordia GR-468-CORE — the de-facto reliability standard for optoelectronic devices, achieved by commercial silicon photonics PDKs (e.g.\ OpenLight on Tower PH18DA, 2025).
\end{itemize}

\section{Master qualification table}
The table on the following pages summarises the qualification flow tailored for a packaged astrophotonic PIC (e.g.\ SiO\textsubscript{2} or Si\textsubscript{3}N\textsubscript{4} AWG with photonic-lantern fibre input feeding a science detector, or an InP/SiPh integrated beam combiner). The 19 steps are grouped from incoming inspection to final acceptance and may be tailored to the mission environment (LEO, L2, GEO, deep space) and to the maturity of the device — see \S\ref{sec:tailoring} (project tailoring matrix) and \S\ref{sec:trl} (TRL roadmap).

\textit{Note: numerical values are typical figures from current astrophotonic test campaigns reported in the literature, the Telcordia GR-468-CORE 2{,}000\,h reliability flow, and ESA/NASA generic specifications. Mission-specific values shall be derived from the project Environmental Specification and the Radiation Environment Specification per ECSS-E-ST-10-04C and NASA-STD-8719.14C.}

\begin{landscape}
\centering\scriptsize
\setlength{\LTpre}{0pt}
\setlength{\LTpost}{0pt}
\begin{longtable}{@{}p{0.5cm} p{3.0cm} p{4.0cm} p{8.0cm} p{6.0cm}@{}}
\rowcolor{esablue}\textcolor{white}{\textbf{\#}} & \textcolor{white}{\textbf{Qualification step}} & \textcolor{white}{\textbf{Applicable standards / test methods}} & \textcolor{white}{\textbf{Technical requirement \& typical specifications}} & \textcolor{white}{\textbf{Test facilities (EU / USA)}}\\
\endfirsthead
\rowcolor{esablue}\textcolor{white}{\textbf{\#}} & \textcolor{white}{\textbf{Qualification step}} & \textcolor{white}{\textbf{Applicable standards / test methods}} & \textcolor{white}{\textbf{Technical requirement \& typical specifications}} & \textcolor{white}{\textbf{Test facilities (EU / USA)}}\\
\endhead

\multicolumn{5}{@{}l}{\cellcolor{section}\textbf{\textcolor{esablue}{A.\ Pre-screening, lot acceptance and baseline characterisation}}}\\

1 & Visual \& dimensional inspection &
\cellbullets{\item MIL-STD-883 M.\,2009 (External visual)\item MIL-STD-883 M.\,2010 (Internal visual)\item ECSS-Q-ST-60C Rev.4 \S6.3\item Telcordia GR-468-CORE \S3.3} &
\cellbullets{\item 100\% inspection of die, waveguide facets, fibre attach, pigtails, package, lid seal, electrical pads.\item No chipping > 25\,\textmu m at facet, no waveguide cracks, no contamination, fibre boot intact.\item Dimensional verification of die, package and pigtail length to drawing.} &
\cellbullets{\item EU: ESA-ESTEC M\&ECL (NL); Tyndall (IE); CSL (BE).\item USA: NASA GSFC Code 562; NASA JPL CE\&A; Aerospace Corp.; NTS.}\\

\rowcolor{rowalt} 2 & Optical / electrical baseline characterisation &
\cellbullets{\item ECSS-Q-ST-70-08C\item ECSS-Q-ST-60-02C / SCC Generic Spec\item Telcordia GR-468-CORE Tables 4-1, 4-2\item IEC 61300 series} &
\cellbullets{\item IL: SiO\textsubscript{2} < 1\,dB/cm, SiN < 0.1\,dB/cm; AWG end-to-end IL $\le$ 5--8\,dB.\item Return loss > 40\,dB; PDL $\le$ 0.3\,dB.\item AWG channel spacing $\le$ 5\% nominal; crosstalk $\le$ -25\,dB.\item Beam combiners: V > 0.95, photometric balance $\pm$5\%, throughput $\ge$ 50\%.\item Birefringence and group delay vs.\ $\lambda$ over operational band (e.g. 600--1700\,nm).} &
\cellbullets{\item EU: ESA OOEL (ESTEC); Fraunhofer IOF (DE); CSEM (CH); CEA-LETI (FR); Astrophotonics/AIP (DE).\item USA: NASA GSFC Code 562; NASA JPL Microdevices Lab; NIST Boulder; MIT Lincoln Lab.}\\

3 & Hermeticity / fine and gross leak &
\cellbullets{\item MIL-STD-883 M.\,1014 (Seal)\item MIL-STD-750 M.\,1071\item ECSS-Q-ST-60-05C\item Telcordia GR-468-CORE \S6.7} &
\cellbullets{\item Fine leak: He bombing Cond.\,A; pass < $5\times10^{-8}$\,atm$\cdot$cm\textsuperscript{3}/s for cavity < 0.05\,cm\textsuperscript{3}.\item Gross leak: fluorocarbon bubble test.\item Applies to packaged InP / SiPh PIC modules with hermetic kovar/butterfly packages.} &
\cellbullets{\item EU: ESA-ESTEC M\&ECL; Alter Technology (ES); Airbus DS (FR/DE).\item USA: NASA GSFC Code 562; Aerospace Corp.; Trace Labs; CORE Systems.}\\

\rowcolor{rowalt} 4 & PIND &
\cellbullets{\item MIL-STD-883 M.\,2020\item ECSS-Q-ST-60-05C} &
\cellbullets{\item 5 PIND cycles at 20\,g (peak), 60--250\,Hz; reject on any noise hit.\item Mandatory for cavity-package PICs (butterfly, BTF, hermetic LCC).} &
\cellbullets{\item EU: ESA-ESTEC M\&ECL; Alter Technology (ES, UK).\item USA: NASA GSFC Code 562; NTS labs; Trace Labs.}\\

\multicolumn{5}{@{}l}{\cellcolor{section}\textbf{\textcolor{esablue}{B.\ Environmental qualification (mechanical, thermal, vacuum, humidity)}}}\\

5 & Sinusoidal \& random vibration &
\cellbullets{\item GSFC-STD-7000B (GEVS) \S2.4\item ECSS-E-ST-10-03C \S5\item MIL-STD-883 M.\,2007\item Telcordia GR-468-CORE \S6.10} &
\cellbullets{\item Sinusoidal qual: 5--100\,Hz, up to 20\,g (axial)/14\,g (lateral), 2\,oct/min, 4 sweeps.\item Random (GEVS qual, components $\le$ 22.7\,kg): 14.1\,g rms, 20--2000\,Hz, 3\,min/axis.\item Acceptance: 10\,g rms, 1\,min/axis.\item Pass: $\Delta$IL $\le$ 0.2\,dB and $\Delta\lambda$ shift $\le$ 10\% channel BW; no fibre-attach failure.} &
\cellbullets{\item EU: ESTEC Test Centre (HYDRA, NL); IABG (DE); CSL (BE); INTA (ES); Airbus DS Toulouse (FR).\item USA: NASA GSFC ETIB; NASA JPL ETL; NASA Glenn Plum Brook; Wyle Labs; NTS.}\\

\rowcolor{rowalt} 6 & Mechanical shock \& constant acceleration &
\cellbullets{\item MIL-STD-883 M.\,2002 (Shock)\item MIL-STD-883 M.\,2001 (Accel.)\item GSFC-STD-7000B \S2.4.5 (Pyroshock)\item ECSS-E-ST-10-03C \S5.6} &
\cellbullets{\item Mech.\ shock: 1500\,g, 0.5\,ms half-sine, 5 axes $\times$ 3 pulses (Cond.\,B); pyroshock SRS to 4000\,g at 10\,kHz.\item Constant acceleration: 5000\,g (Cond.\,D) or 20\,000\,g (Cond.\,E) for hermetic die-package interface.\item Pass: no facet damage, $\Delta$IL $\le$ 0.5\,dB.} &
\cellbullets{\item EU: ESTEC Test Centre (NL); IABG (DE); CSL (BE); ONERA (FR).\item USA: NASA GSFC ETIB; NASA JPL ETL; NASA MSFC; Wyle/Element labs.}\\

7 & Temperature cycling &
\cellbullets{\item MIL-STD-883 M.\,1010\item MIL-STD-883 M.\,1011 (Thermal shock)\item ECSS-Q-ST-70-04C\item Telcordia GR-468-CORE \S6.4} &
\cellbullets{\item Qual: -55\,\textcelsius\ / +125\,\textcelsius\ (Cond.\,C), 500 cycles, 10\,min dwell, < 10\,\textcelsius/min ramp.\item Astrophotonic-tailored: -40\,\textcelsius\ / +85\,\textcelsius\ (Cond.\,B), 100--500 cycles for SiO\textsubscript{2}/SiN PICs and ULI waveguides.\item Beam combiner spec: phase stability < $\lambda$/100 over orbital cycle.\item Pass: $\Delta$IL $\le$ 0.5\,dB, $\Delta$crosstalk $\le$ 2\,dB.} &
\cellbullets{\item EU: ESTEC M\&ECL \& Test Centre; CSL (BE); Airbus DS; Fraunhofer IZM (DE); CSEM (CH).\item USA: NASA GSFC Code 562 \& ETIB; NASA JPL ETL; NASA JSC; Aerospace Corp.; NTS.}\\

\rowcolor{rowalt} 8 & Thermal vacuum (TVAC) cycling &
\cellbullets{\item GSFC-STD-7000B \S2.6.2\item ECSS-Q-ST-70-04C\item ECSS-E-ST-10-03C \S5.4} &
\cellbullets{\item Qual: 4 cycles min.\ (8 recommended for new tech); pressure $\le$ $1\times10^{-5}$\,Torr.\item Hot/cold soak: $T_{\max}$+10\,\textcelsius\ / $T_{\min}$-10\,\textcelsius\ of mission predict.\item Functional + performance test at each plateau (in-vacuum optical IL, AWG channel positions, beam-combiner null depth).\item Outgassing: TML < 1.0\%, CVCM < 0.1\% per ECSS-Q-ST-70-02C / ASTM E595.} &
\cellbullets{\item EU: ESTEC LSS \& Phenix-VTC (NL); CSL FOCAL \& VTC (BE); IABG (DE); INTA (ES).\item USA: NASA GSFC SES Bldg 7/10; NASA JPL 25-ft \& 10-ft Space Simulators; NASA JSC Chambers A \& B; NASA Glenn SPF Plum Brook.}\\

9 & Damp heat / humidity bias &
\cellbullets{\item MIL-STD-883 M.\,1004\item Telcordia GR-468-CORE \S6.5 (THB)\item IEC 60068-2-67} &
\cellbullets{\item Temperature-humidity bias (THB): 85\,\textcelsius\ / 85\,\%RH, 1000\,h, with operational bias on active PICs.\item Damp heat storage (THS): 85\,\textcelsius\ / 85\,\%RH, 1000\,h, unbiased.\item Critical for non-hermetic SiPh / InP heterogeneously integrated PICs.\item Pass: $\Delta$IL $\le$ 0.5\,dB, $\Delta I_\text{th}\le$ 10\%, $\Delta$responsivity $\le$ 10\%.} &
\cellbullets{\item EU: ESA-ESTEC M\&ECL; Alter Technology (ES); Tyndall (IE); Fraunhofer IZM (DE).\item USA: NASA GSFC Code 562; Aerospace Corp.; NTS; CORE Labs.}\\

\rowcolor{rowalt} 10 & Outgassing / contamination &
\cellbullets{\item ECSS-Q-ST-70-02C\item ASTM E595\item NASA-STD-6016 \S4.4} &
\cellbullets{\item Thermal vacuum outgassing: 125\,\textcelsius, 24\,h, $10^{-5}$\,Torr.\item Limits: TML $\le$ 1.0\%, CVCM $\le$ 0.1\%, WVR $\le$ 1.0\%.\item Critical for adhesives at fibre-to-chip interfaces (UV-cure epoxies, EpoTek 353ND etc.) and PIC encapsulants.} &
\cellbullets{\item EU: ESA-ESTEC M\&ECL (NL, ISO/IEC 17025); CNES Toulouse (FR); DLR Cologne (DE); AIT Seibersdorf (AT).\item USA: NASA GSFC Contamination \& Coatings Lab (Code 546); NASA JPL Contamination Control.}\\

\multicolumn{5}{@{}l}{\cellcolor{section}\textbf{\textcolor{esablue}{C.\ Reliability and life testing}}}\\

11 & High-Temperature Operating Life (HTOL) &
\cellbullets{\item MIL-STD-883 M.\,1005\item Telcordia GR-468-CORE \S6.2\item JEDEC JESD22-A108} &
\cellbullets{\item Active PICs (lasers, SOAs, modulators, PDs): 2000\,h at $T_{j,\max}$ (typ.\ 85\,\textcelsius) with nominal bias.\item Acceleration via Arrhenius ($E_a$ 0.4--0.7\,eV for III-V lasers).\item Sample: $\ge$ 22 devices, 0 failures (LTPD 10).\item Pass: $\Delta I_\text{th}\le$ 20\%, slope eff.\ drift $\le$ 20\%, $V_\pi$ drift $\le$ 10\%.} &
\cellbullets{\item EU: ESA-ESTEC M\&ECL; Alter Technology (ES); Airbus DS; Tyndall (IE).\item USA: NASA GSFC Code 562; Aerospace Corp.; NIST; OEM in-house (Coherent, II-VI/Lumentum).}\\

\rowcolor{rowalt} 12 & Burn-in &
\cellbullets{\item MIL-STD-883 M.\,1015\item ECSS-Q-ST-60-05C \S6.7\item Telcordia GR-468-CORE \S5} &
\cellbullets{\item Static / dynamic burn-in 168\,h at 125\,\textcelsius\ (Cond.\,A/B) for active components.\item Purpose: precipitate infant mortality; reject on > 20\% drift in optical output power, threshold or wavelength.} &
\cellbullets{\item EU: ESA-ESTEC M\&ECL; Alter Technology (ES); Tyndall (IE).\item USA: NASA GSFC Code 562; Aerospace Corp.; NTS; Trace Labs.}\\

\multicolumn{5}{@{}l}{\cellcolor{section}\textbf{\textcolor{esablue}{D.\ Radiation hardness assurance (RHA)}}}\\

13 & Total Ionising Dose (TID) — gamma &
\cellbullets{\item ESCC Basic Spec.\ 22900\item MIL-STD-883 M.\,1019\item ECSS-Q-ST-60-15C Rev.1 (Mar 2025)\item ASTM F1892} &
\cellbullets{\item Co-60 gamma exposure to 30--300\,krad(Si) for LEO/GEO; up to 1\,Mrad(Si) for Jovian/JUICE-class.\item Dose rate 36--360\,rad/h (low) and 0.5--50\,krad/h (standard); both required for ELDRS-sensitive parts.\item Pass for passive PICs (SiO\textsubscript{2}/SiN AWG, ULI waveguides): $\Delta$IL < 0.5\,dB up to 1\,Mrad — typically pass with margin.\item Active PICs: track $I_\text{th}$, $I_d$, responsivity, $V_\pi$, optical power vs.\ dose.} &
\cellbullets{\item EU: ESA-ESTEC Co-60 (NL); UCL Cyclotron GIF (BE); RADEF (FI); Fraunhofer INT (DE); IRSN (FR).\item USA: NASA GSFC REAG; NASA JPL Co-60 + IBA Dynamitron; Sandia GIF; NRL; ARL.}\\

\rowcolor{rowalt} 14 & Displacement Damage Dose (DDD/TNID) &
\cellbullets{\item ESCC Basic Spec.\ 25100\item ECSS-Q-ST-60-15C Rev.1\item MIL-STD-750 M.\,1017\item ASTM E722 (NIEL)} &
\cellbullets{\item Proton fluence: 50--200\,MeV protons to $1\times10^{11}$--$1\times10^{12}$\,p/cm\textsuperscript{2} (LEO) up to $1\times10^{13}$\,p/cm\textsuperscript{2} (Europa Clipper-class).\item DDD measured via NIEL scaling; primary concern for InP / Ge photodiodes (dark current rise) and SOAs.\item Recent results (laser-written SiO\textsubscript{2} waveguides): negligible IL change after $6\times10^{10}$\,p/cm\textsuperscript{2} at 60\,MeV.\item Pass: dark current rise $\le$ 2$\times$ initial; responsivity loss $\le$ 10\%.} &
\cellbullets{\item EU: UCL CRC LIF (BE) protons 10--75\,MeV; PSI PIF (CH) 6--230\,MeV; KVI-CART (NL); RADEF (FI); CNA Sevilla (ES).\item USA: NASA NSRL @ BNL; UC Davis Crocker; MGH Burr Center; Indiana Univ.; Texas A\&M; Loma Linda.}\\

15 & Single Event Effects (SEE) &
\cellbullets{\item ESCC Basic Spec.\ 25100\item ECSS-Q-ST-60-15C Rev.1\item JEDEC JESD57\item MIL-STD-750 M.\,1080} &
\cellbullets{\item Heavy-ion LET range 1--80\,MeV$\cdot$cm\textsuperscript{2}/mg; fluence $1\times10^{7}$\,ions/cm\textsuperscript{2} per LET.\item Targets: SEU (digital control of PIC heaters/MZIs), SET in photodiodes/TIAs, SEL/SEB in driver electronics.\item LET threshold $\ge$ 37\,MeV$\cdot$cm\textsuperscript{2}/mg for SEL-immune classification.\item Pulsed-laser SEE: 1064\,nm sub-ps, complementary screening for sensitive nodes.} &
\cellbullets{\item EU: UCL CRC HIF (BE) -- CYCLONE; GANIL (FR); JYFL/RADEF (FI); Legnaro INFN (IT); MBDA SEREEL (UK).\item USA: Texas A\&M Cyclotron; LBNL 88-Inch Cyclotron; Brookhaven NSRL; Michigan State NSCL/FRIB; NRL pulsed-laser; Vanderbilt.}\\

\rowcolor{rowalt} 16 & UV / particle exposure for photonic materials &
\cellbullets{\item ECSS-Q-ST-70-06C\item ASTM E512} &
\cellbullets{\item Combined VUV (115--200\,nm) + solar UV (200--400\,nm) up to 2000 ESH.\item Particularly relevant for unprotected silica/SiN waveguide cladding and polymer-based PICs.\item Pass: cladding loss increase $\le$ 0.1\,dB/cm at operating $\lambda$; no surface darkening.} &
\cellbullets{\item EU: ESA-ESTEC Materials Lab; CSL (BE); ONERA (FR); DLR (DE); AIT Seibersdorf (AT).\item USA: NASA GSFC Materials Engineering; NASA JPL M\&P; Aerospace Corp.}\\

\multicolumn{5}{@{}l}{\cellcolor{section}\textbf{\textcolor{esablue}{E.\ ESD, EMC and final acceptance}}}\\

17 & ESD sensitivity &
\cellbullets{\item MIL-STD-883 M.\,3015 (HBM)\item JEDEC JS-001 (HBM), JS-002 (CDM)\item Telcordia GR-468-CORE \S6.11} &
\cellbullets{\item HBM: minimum class 1A (250\,V) for III-V active PICs; class 2 (2\,kV) preferred.\item CDM: $\ge$ 250\,V.\item Mandatory before integration into spacecraft optical payload.} &
\cellbullets{\item EU: ESA-ESTEC M\&ECL; Tyndall (IE); Alter Technology (ES); Fraunhofer IZM (DE).\item USA: NASA GSFC Code 562; Aerospace Corp.; NTS; CORE Labs.}\\

\rowcolor{rowalt} 18 & EMC / EMI at instrument level &
\cellbullets{\item GSFC-STD-7000B \S2.5 (tailored MIL-STD-461)\item MIL-STD-461G\item ECSS-E-ST-20-07C} &
\cellbullets{\item Conducted/radiated emissions and susceptibility testing of full PIC-based instrument.\item CE101/CE102, RE101/RE102, CS101/CS114/CS115/CS116, RS101/RS103.\item Magnetic cleanliness for missions like LISA / Swarm / Plato.} &
\cellbullets{\item EU: ESTEC EMC Lab (NL); IABG MFSA (DE); Airbus EMC labs; INTA (ES).\item USA: NASA GSFC EMC Lab; NASA JPL EMC Lab; NTS; Wyle/Element.}\\

19 & End-of-line / final acceptance test &
\cellbullets{\item ECSS-Q-ST-10-09C\item ECSS-Q-ST-20C\item NASA-STD-8739.11 / EEE-INST-002} &
\cellbullets{\item Repeat baseline optical/electrical characterisation; compare to pre-environmental data.\item $\Delta$ criteria: $\Delta$IL $\le$ 1\,dB (passive PIC), $\Delta\lambda$ shift $\le$ 50\,pm, beam-combiner visibility loss $\le$ 5\,pp.\item Generation of full Qualification Test Report and Declared Materials/Components Lists.} &
\cellbullets{\item EU: ESTEC OOEL \& M\&ECL; CSL (BE); Alter Technology (ES); ESA SCC certification authority.\item USA: NASA GSFC Photonics Group (Code 562); NASA JPL CE\&A; NEPP.}\\

\end{longtable}
\end{landscape}

\section{Project-specific tailoring matrix: HWO L2 vs.\ LEO smallsat}\label{sec:tailoring}
Two reference profiles bound the practical application of the master qualification flow to astrophotonic PIC payloads:

\paragraph{LEO smallsat demonstrator.} A 6U--12U CubeSat or microsatellite hosting a PIC-based spectrograph or beam combiner as a technology demonstration. Typical orbit 400--600\,km, sun-synchronous, 1--3 year lifetime, NASA Class~D / ESA Category~4 mission classification. Total ionising dose mostly from trapped electrons and protons in the inner belt and the South Atlantic Anomaly; thermal cycling driven by $\sim$16 eclipse cycles per day. Tolerant of single-string architectures, COTS-with-uprating, and commercial reliability data.

\paragraph{HWO-class L2 flagship mission.} The Habitable Worlds Observatory and similar UV/optical/IR flagships are designed for a halo orbit at the Earth--Sun L2 point, $1.5\times10^6$\,km from Earth, with a mission lifetime of 5--10+ years and a planned servicing capability. NASA Class~A mission classification. The radiation environment is dominated by galactic cosmic rays (GCR) and solar particle events (SPE), with no trapped-belt component but a harder, more penetrating spectrum. Thermally the environment is far more stable than LEO, but the science requirements ($10^{-10}$ raw contrast for direct exoplanet imaging) impose sub-mK temperature and sub-pm phase stability on the optical bench.

The matrix below maps each of the 19 qualification steps of \S\ref{sec:tailoring} onto these two reference profiles.

\begin{landscape}
\centering\scriptsize
\setlength{\LTpre}{0pt}\setlength{\LTpost}{0pt}
\begin{longtable}{@{}p{0.4cm} p{2.6cm} p{5.5cm} p{5.5cm} p{6.0cm}@{}}
\rowcolor{esablue}\textcolor{white}{\textbf{\#}} & \textcolor{white}{\textbf{Qualification step}} &
\cellcolor{leohdr}\textcolor{white}{\textbf{LEO smallsat demonstrator}} &
\cellcolor{hwohdr}\textcolor{white}{\textbf{HWO-class flagship (L2)}} &
\textcolor{white}{\textbf{Tailoring rationale}}\\
\endfirsthead
\rowcolor{esablue}\textcolor{white}{\textbf{\#}} & \textcolor{white}{\textbf{Qualification step}} &
\cellcolor{leohdr}\textcolor{white}{\textbf{LEO smallsat demonstrator}} &
\cellcolor{hwohdr}\textcolor{white}{\textbf{HWO-class flagship (L2)}} &
\textcolor{white}{\textbf{Tailoring rationale}}\\
\endhead

1 & Visual \& dimensional inspection &
\cellcolor{leo} Sample-based (AQL 1.0 or 25\%); waveguide facets only on flight units. &
\cellcolor{hwo} 100\% inspection of every chip, fibre attach, package; 25\,\textmu m chip-edge limit; SEM on representative facets. &
Flight class S/Class 1 vs Class 3 part screening: HWO follows EEE-INST-002 / NASA-STD-8739.11 Level 1 (full); LEO demos may use Level 3 / PAL 4 commercial.\\

\rowcolor{rowalt} 2 & Optical / electrical baseline &
\cellcolor{leo} Operating-band IL, PDL, AWG/MZI spectral response at 1\,T (room temp); 5--10 device sample. &
\cellcolor{hwo} Full operating range (UV--NIR, 200--1700\,nm), 3\,T (cold/ambient/hot), birefringence map, beam-combiner V \& nulling depth, photometric stability < 0.1\% over 24\,h. &
HWO contrast $10^{-10}$ requires sub-mK and sub-pm phase stability — characterisation must encompass operational dynamic range. LEO demos validate basic functionality only.\\

3 & Hermeticity & 
\cellcolor{leo} Optional for non-hermetic SiPh demonstrator with conformal coating. If hermetic: He fine leak at $5\times10^{-7}$\,atm$\cdot$cm\textsuperscript{3}/s. &
\cellcolor{hwo} Mandatory hermetic packaging for active PICs ($5\times10^{-8}$\,atm$\cdot$cm\textsuperscript{3}/s). Long-term hermeticity verified via residual gas analysis (RGA) per MIL-STD-883 M.\,1018. &
5--10\,yr mission with possible servicing requires demonstrated hermetic seal stability over the full operational life; condensation on chip facets at L2 cold side would degrade contrast.\\

\rowcolor{rowalt} 4 & PIND &
\cellcolor{leo} Waiver acceptable for non-cavity packaged PICs (epoxy-encapsulated SiPh). &
\cellcolor{hwo} Mandatory; tighter rejection threshold (no anomalies in 5 cycles). &
L2 missions cannot be repaired beyond robotic servicing — single particle on chip facet is mission-critical.\\

5 & Sinusoidal \& random vibration &
\cellcolor{leo} Random 10\,g rms, 20--2000\,Hz, 1\,min/axis (acceptance only); Falcon 9 / Electron / RFA-class envelope. &
\cellcolor{hwo} Random 14.1\,g rms, 20--2000\,Hz, 3\,min/axis (qual); Sin.\ 20\,g axial / 14\,g lateral; Starship / SLS / New Glenn envelope tailored to launcher. &
Both follow GEVS \S2.4; HWO requires full qualification + protoflight; LEO demos can use protoflight or acceptance only depending on heritage of carrier bus.\\

\rowcolor{rowalt} 6 & Mechanical shock \& constant accel. &
\cellcolor{leo} Mech.\ shock 500--1000\,g half-sine, 0.5\,ms; pyroshock SRS to 2000\,g if separation device present. &
\cellcolor{hwo} Mech.\ shock 1500\,g, 0.5\,ms (Cond.\,B); pyroshock SRS 4000\,g at 10\,kHz; constant acceleration 5000--20\,000\,g for hermetic die-package interface. &
Larger launcher and more separation events (fairing, SCs deployment, possible robotic servicing) drive higher shock spectra for HWO.\\

7 & Temperature cycling (component) &
\cellcolor{leo} -40\,\textcelsius\ / +85\,\textcelsius, 100 cycles (matching LEO eclipse cycling, $\sim$16 cycles/day $\times$ $\sim$3\,yr $\approx$ 17\,500 cycles, derated by analysis). &
\cellcolor{hwo} -55\,\textcelsius\ / +125\,\textcelsius, 500 cycles (Cond.\,C); plus dedicated cryogenic cycling 80\,K $\rightarrow$ 300\,K, 50 cycles, for instrument-bench PICs. &
L2 has very stable thermal environment but instruments can run cold (e.g.\ 80\,K for IR PICs); the high LEO cycle count is partly addressed by accelerated testing using larger $\Delta T$.\\

\rowcolor{rowalt} 8 & Thermal vacuum (TVAC) cycling &
\cellcolor{leo} 4 cycles, $\Delta T = T_{\max}{+}5$\,\textcelsius\ / $T_{\min}{-}5$\,\textcelsius, $P \le 1\times10^{-5}$\,Torr; functional + IL at each plateau. &
\cellcolor{hwo} 8 cycles min., $\Delta T = T_{\max}{+}10$\,\textcelsius\ / $T_{\min}{-}10$\,\textcelsius; instrument-level TVAC at 80\,K with optical pickoff for AWG/beam-combiner test; sub-\textmu K stability demonstration. &
HWO contrast budget allocates only a few ppm to thermal drift — stability rather than range is the driver. Specific cryogenic chambers needed (CSL FOCAL, NASA GSFC SES, JPL Cryo).\\

9 & Damp heat / humidity bias &
\cellcolor{leo} Optional; 85\,\textcelsius / 85\,\%RH, 168--500\,h on non-hermetic devices. &
\cellcolor{hwo} Mandatory 1000\,h at 85\,\textcelsius / 85\,\%RH on representative non-hermetic builds; not required if all PICs are hermetic. &
Long shelf life and integration period (up to 5\,yr ground storage before launch) drives damp-heat exposure for HWO.\\

\rowcolor{rowalt} 10 & Outgassing / contamination &
\cellcolor{leo} ECSS-Q-ST-70-02C: TML $\le$ 1.0\%, CVCM $\le$ 0.1\% (lot acceptance). &
\cellcolor{hwo} Same TML/CVCM limits, plus VCM $\le 1\times10^{-14}$\,g/cm\textsuperscript{2}/s for adhesives near coronagraph and PIC facets; certified to ASTM E1559. &
$10^{-10}$ contrast precludes any condensable on optical surfaces — contamination budget is $\sim$1\,nm equivalent thickness over mission.\\

11 & HTOL &
\cellcolor{leo} Telcordia 1000\,h at $T_{j,\max}$ with 5--10 sample (LTPD 20). &
\cellcolor{hwo} Telcordia 2000\,h at $T_{j,\max}$ + Arrhenius extrapolation to 87\,600\,h (10\,yr) with $E_a > 0.5$\,eV; 22-device sample (LTPD 10); 0 failures required. &
ESA/NASA flight qualification requires demonstrated reliability at full mission lifetime including derating; LEO demos use shorter-time confidence with FIT projection.\\

\rowcolor{rowalt} 12 & Burn-in &
\cellcolor{leo} 168\,h at 125\,\textcelsius, 100\,\% units. &
\cellcolor{hwo} 168\,h at 125\,\textcelsius\ + 24\,h at $T_{j,\max}$ with optical/electrical functional monitoring (intermittent failures detected). &
Infant-mortality rejection critical for non-serviceable lifetime; HWO adds in-situ functional monitoring during burn-in.\\

13 & Total Ionising Dose (TID) &
\cellcolor{leo} Mission-end TID (4\,mm Al equiv.): $\sim$3--5\,krad(Si) for 1\,yr, 10--15\,krad(Si) for 3\,yr at 500\,km. Test to 30\,krad(Si) with RDM = 2. &
\cellcolor{hwo} Mission-end TID (5\,mm Al equiv.): $\sim$2--5\,krad(Si)/yr at L2 (no trapped belts; GCR + solar). Test to 50--100\,krad(Si) for 10\,yr with RDM = 2 plus solar-event allocation. &
L2 TID is dominated by GCR + solar protons — typically lower than LEO inside SAA per year, but higher cumulative for HWO due to lifetime ($\times$3--5).\\

\rowcolor{rowalt} 14 & DDD &
\cellcolor{leo} Proton fluence (E > 50\,MeV, 4\,mm Al): $\sim 2\times10^{9}$\,p/cm\textsuperscript{2}/yr in LEO. Test to $1\times10^{10}$\,p/cm\textsuperscript{2} at 60\,MeV. &
\cellcolor{hwo} Proton fluence at L2 (5\,mm Al): GCR + solar event; $\sim 5\times10^{10}$\,p/cm\textsuperscript{2} for 10\,yr with worst-case SPE allocation. Test to $1\times10^{11}$\,p/cm\textsuperscript{2} at 50, 100, 200\,MeV. &
GCR spectrum is harder than trapped belts $\rightarrow$ multi-energy DDD test needed; particularly important for InP/Ge photodiodes whose dark current scales with NIEL$\times$fluence.\\

15 & SEE (heavy ions) &
\cellcolor{leo} Test to LET 37\,MeV$\cdot$cm\textsuperscript{2}/mg; SEU rate budget acceptable up to $10^{-5}$\,/device/day. &
\cellcolor{hwo} Test to LET 80\,MeV$\cdot$cm\textsuperscript{2}/mg for SEL-immunity; SEU rate < $10^{-7}$\,/device/day for safety-critical functions; SEFI tolerated only with autonomous recovery. &
Outside Earth's magnetosphere GCR LET spectrum extends higher; HWO autonomy and 5--10+\,yr lifetime require strict SEFI / SEL classification.\\

\rowcolor{rowalt} 16 & UV / particle on materials &
\cellcolor{leo} UV: 100 ESH (1\,ESH $\approx$ 1 sun-hour) for 1--3\,yr LEO operation behind sun shield. &
\cellcolor{hwo} UV: 1000--2000 ESH for 10\,yr at L2 with possible direct-sun exposure during attitude excursions; VUV (115--200\,nm) test mandatory for UV-band PICs. &
Direct UV exposure in HWO bandpass (down to 100\,nm) would damage unprotected polymer claddings or organic adhesives near PIC facets.\\

17 & ESD &
\cellcolor{leo} HBM $\ge$ 250\,V (Class 1A) for active components. &
\cellcolor{hwo} HBM $\ge$ 2\,kV (Class 2) and CDM $\ge$ 500\,V for all flight units; full ESD-control program during AIT. &
Higher voltage class required by NASA GSFC Code 562 / ESA SCC for flight-grade active photonics; LEO demos can accept Class 1A.\\

\rowcolor{rowalt} 18 & EMC at instrument level &
\cellcolor{leo} MIL-STD-461 tailored RE/CE only; CS for primary power and signal. &
\cellcolor{hwo} Full GEVS \S2.5 EMC suite + magnetic cleanliness for L2 fine-pointing; LISA-class magnetic budgeting if PIC drivers near coronagraph. &
HWO fine-pointing and contrast stability impose magnetic cleanliness requirements absent in routine LEO smallsats.\\

19 & End-of-line acceptance &
\cellcolor{leo} $\Delta$-criteria documented vs.\ baseline; project-specific waivers permitted. &
\cellcolor{hwo} Strict $\Delta$-criteria + flight Test Readiness Review + Independent Review Board sign-off; full traceability to ECSS-Q-ST-10/20 + NPR 8735.1. &
Programmatic rigor of NASA flagship vs.\ Class~D smallsat; documentation hierarchy is one of the largest cost differentials.\\

\end{longtable}
\end{landscape}

\section{TRL-vs-test-coverage roadmap}\label{sec:trl}
\begin{figure}[htbp]
\centering
\includegraphics[width=0.86\linewidth]{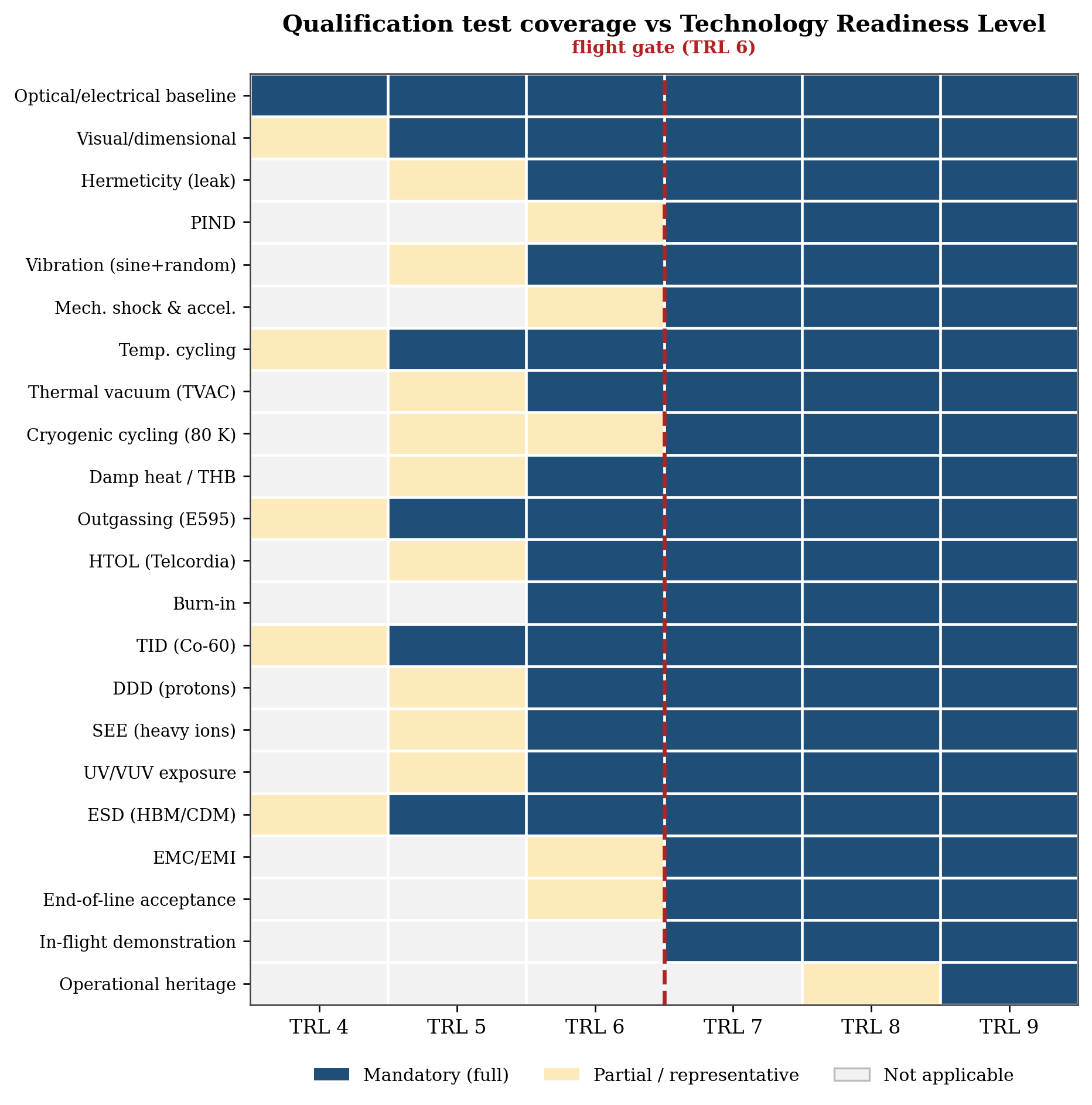}
\caption{Graphical form of the TRL-vs-test-coverage matrix of \S\ref{sec:trl}: for each qualification activity, the cell shading indicates whether the test is mandatory (full), partial/representative, or not applicable at a given technology-readiness level. The dashed line marks the TRL-6 flight-readiness gate. Test methods and governing standards are those tabulated in the master qualification flow of \S\ref{sec:tailoring}.}
\label{fig:trlcoverage}
\end{figure}

Both NASA (NPR 7123.1, NPR 7120.8) and ESA (ECSS-E-AS-11C, adopted from ISO 16290) use a 1--9 Technology Readiness Level scale to track maturation. For astrophotonic PICs the typical entry point is TRL 3 (analytical/experimental proof of concept) and the gate to flight is TRL 6 (system demonstration in a relevant environment). The matrix below shows which qualification activities of \S\ref{sec:tailoring} are typically required at each TRL and which review or mission gate they support. Figure~\ref{fig:trlcoverage} presents the same coverage map graphically.

\paragraph{Coding.} \trlfull\ = mandatory at this TRL\quad \trlpart\ = partial (representative samples or relaxed conditions)\quad \trlnone\ = not applicable. Mission gates: PDR = Preliminary Design Review; CDR = Critical Design Review; QR = Qualification Review; MIP = Mandatory Inspection Point; AR = Acceptance Review; FAR = Flight Acceptance Review; EOL = End-of-life heritage report.

\subsection{Coverage matrix}

\begin{landscape}
\centering\footnotesize
\setlength{\LTpre}{0pt}\setlength{\LTpost}{0pt}
\begin{longtable}{@{}p{6.5cm} C{1.6cm} C{1.6cm} C{1.6cm} C{1.6cm} C{1.6cm} C{1.6cm} C{2.0cm}@{}}
\rowcolor{esablue}\textcolor{white}{\textbf{Test category}} & \textcolor{white}{\textbf{TRL 4}} & \textcolor{white}{\textbf{TRL 5}} & \textcolor{white}{\textbf{TRL 6}} & \textcolor{white}{\textbf{TRL 7}} & \textcolor{white}{\textbf{TRL 8}} & \textcolor{white}{\textbf{TRL 9}} & \textcolor{white}{\textbf{Mission gate}}\\
\endfirsthead
\rowcolor{esablue}\textcolor{white}{\textbf{Test category}} & \textcolor{white}{\textbf{TRL 4}} & \textcolor{white}{\textbf{TRL 5}} & \textcolor{white}{\textbf{TRL 6}} & \textcolor{white}{\textbf{TRL 7}} & \textcolor{white}{\textbf{TRL 8}} & \textcolor{white}{\textbf{TRL 9}} & \textcolor{white}{\textbf{Mission gate}}\\
\endhead

Optical / electrical baseline characterisation & \trlfull & \trlfull & \trlfull & \trlfull & \trlfull & \trlfull & PDR/CDR\\
\rowcolor{rowalt} Visual / dimensional inspection & \trlpart & \trlfull & \trlfull & \trlfull & \trlfull & \trlfull & MIP\\
Hermeticity (fine + gross leak) & \trlnone & \trlpart & \trlfull & \trlfull & \trlfull & \trlfull & CDR\\
\rowcolor{rowalt} PIND & \trlnone & \trlnone & \trlpart & \trlfull & \trlfull & \trlfull & AIT\\
Sinusoidal \& random vibration & \trlnone & \trlpart & \trlfull & \trlfull & \trlfull & \trlfull & QR\\
\rowcolor{rowalt} Mechanical shock \& constant acceleration & \trlnone & \trlnone & \trlpart & \trlfull & \trlfull & \trlfull & QR\\
Temperature cycling (atm.) & \trlpart & \trlfull & \trlfull & \trlfull & \trlfull & \trlfull & PDR/CDR\\
\rowcolor{rowalt} Thermal vacuum (TVAC) cycling & \trlnone & \trlpart & \trlfull & \trlfull & \trlfull & \trlfull & QR\\
Cryogenic cycling (80\,K) & \trlnone & \trlpart & \trlhalf & \trlfull & \trlfull & \trlfull & QR\\
\rowcolor{rowalt} Damp heat / THB & \trlnone & \trlpart & \trlfull & \trlfull & \trlfull & \trlfull & QR\\
Outgassing (ASTM E595 / TML, CVCM) & \trlpart & \trlfull & \trlfull & \trlfull & \trlfull & \trlfull & PDR\\
\rowcolor{rowalt} HTOL (Telcordia 2000\,h) & \trlnone & \trlpart & \trlfull & \trlfull & \trlfull & \trlfull & QR\\
Burn-in & \trlnone & \trlnone & \trlfull & \trlfull & \trlfull & \trlfull & AR\\
\rowcolor{rowalt} TID (Co-60 gamma) & \trlpart & \trlfull & \trlfull & \trlfull & \trlfull & \trlfull & QR\\
DDD (proton irradiation) & \trlnone & \trlpart & \trlfull & \trlfull & \trlfull & \trlfull & QR\\
\rowcolor{rowalt} SEE (heavy ions / pulsed laser) & \trlnone & \trlpart & \trlfull & \trlfull & \trlfull & \trlfull & QR\\
UV/VUV exposure (ECSS-Q-ST-70-06C) & \trlnone & \trlpart & \trlfull & \trlfull & \trlfull & \trlfull & QR\\
\rowcolor{rowalt} ESD (HBM / CDM) & \trlpart & \trlfull & \trlfull & \trlfull & \trlfull & \trlfull & AR\\
EMC / EMI (instrument level) & \trlnone & \trlnone & \trlpart & \trlfull & \trlfull & \trlfull & QR\\
\rowcolor{rowalt} End-of-line acceptance \& test report & \trlnone & \trlnone & \trlpart & \trlfull & \trlfull & \trlfull & AR/FAR\\
In-flight demonstration & \trlnone & \trlnone & \trlnone & \trlfull & \trlfull & \trlfull & Launch\\
\rowcolor{rowalt} Operational mission heritage & \trlnone & \trlnone & \trlnone & \trlnone & \trlpart & \trlfull & EOL\\

\end{longtable}
\end{landscape}

\subsection{Reading the matrix}
A laboratory PIC AWG that has demonstrated room-temperature spectral response and basic temperature cycling sits at \textbf{TRL 4}. Once the same chip has been tested in a representative thermal-vacuum environment with the optical interface (fibre array unit, mounting bench) it reaches \textbf{TRL 5}. A full instrument prototype that has passed the qualification flow (vibration, TVAC, radiation, EMC) at the relevant levels, including representative cryogenic operation for IR spectrographs, reaches \textbf{TRL 6} — the typical entry level for inclusion in a flight project per NASA SP-20205003605 (TRA Best Practices Guide). \textbf{TRL 7} requires demonstration of the prototype in space (typically via a smallsat technology demonstrator like the LEO profile in \S\ref{sec:tailoring}), \textbf{TRL 8} requires the actual flight unit to be qualified, and \textbf{TRL 9} requires successful operational use in the same environment.

For HWO-class missions, NASA Strategic Astrophysics Technology (SAT) and ESA TRP/GSTP funding instruments typically require advancement to TRL 5--6 before component selection at the mission System Requirements Review (SRR), and TRL 6 prior to Preliminary Design Review (PDR). LEO smallsat technology demonstrators are themselves often the vehicle by which a TRL-5 PIC reaches TRL 7.

\section{Notes on tailoring for astrophotonic PICs}
Astrophotonic devices used in space spectrographs and beam combiners differ from telecom-grade PICs in three important respects, all of which affect the qualification approach:

\begin{itemize}[leftmargin=1.4em]
  \item \textbf{Wavelength range.} From visible ($\approx 500$\,nm) for HARPS/ESPRESSO-class radial-velocity PICs up to mid-IR (3--10\,\textmu m) for nulling interferometers, well beyond the 1310/1550\,nm telecom band for which Telcordia GR-468 was originally written. Optical pre/post characterisation must therefore be performed at the actual operational wavelength.
  \item \textbf{Photometric and phase requirements.} Precision spectrographs require throughput stability better than 0.1\,\% over an orbit; beam combiners (ABCD, DBC, nullers) require visibility stability and phase stability $< \lambda/100$. These tolerances are tighter than typical telecom IL drift acceptance and drive bespoke thermal-vacuum performance verifications at the instrument level.
  \item \textbf{Cryogenic operation.} For IR astrophotonic spectrographs the operational temperature can be 80\,K or below, requiring specific cryogenic cycling and test capabilities (CSL FOCAL chamber, NASA GSFC SES, NASA JPL Cryo Test Facility, NIST Boulder).
  \item \textbf{Radiation.} Passive silica/Si\textsubscript{3}N\textsubscript{4} waveguides and ULI-written waveguides have shown excellent radiation tolerance (no measurable IL change up to 1\,Mrad TID and $10^{11}$\,p/cm\textsuperscript{2}). The dominant radiation concerns are therefore the active III-V building blocks (lasers, SOAs, photodiodes) and the supporting ASIC/FPGA control electronics.
\end{itemize}

\section{PIC material platforms surveyed: UV, visible and NIR}\label{sec:platforms}
The qualification flow of \S\ref{sec:tailoring}--\S\ref{sec:trl} is in principle platform-agnostic, but in practice the choice of waveguide material drives much of the technology-readiness and radiation-tolerance budget — and therefore the test campaign emphasis. This section surveys the material platforms most relevant to astrophotonic and space-photonic instruments, ordered by spectral coverage. The match between platform and mission is determined first by wavelength, second by radiation tolerance, and third by packaging maturity and active-device availability (Terrasanta et al.~\cite{terrasanta2025}; Blumenthal et al.~\cite{blumenthal2020}).

Three rapidly growing space application classes are pulling PIC technology into flight hardware in parallel: (i) \emph{optical satellite communications} (NASA LCRD in GEO since 2021, TBIRD 200\,Gbps CubeSat downlinks in 2022, DSOC at Psyche in 2023) (Schieler et al.~\cite{schieler2023}; Biswas et al.~\cite{biswas2024dsoc}; Terrasanta et al.~\cite{terrasanta2025}); (ii) \emph{astronomical instrumentation} where the 2023 Astrophotonics Roadmap identifies PICs as a key enabler for next-generation ELTs and HWO (Jovanovic et al.~\cite{jovanovic2023,madhav2024}); and (iii) \emph{Earth-observation and proximity-operations LiDAR} where FMCW coherent architectures are converging on integrated photonic transmitters and receivers (Martin et al.~\cite{martin2018}; Lihachev et al.~\cite{lihachev2024}).

\subsection{Material platform landscape}\label{sec:platforms_overview}

\begin{figure}[htbp]
\centering
\includegraphics[width=\linewidth]{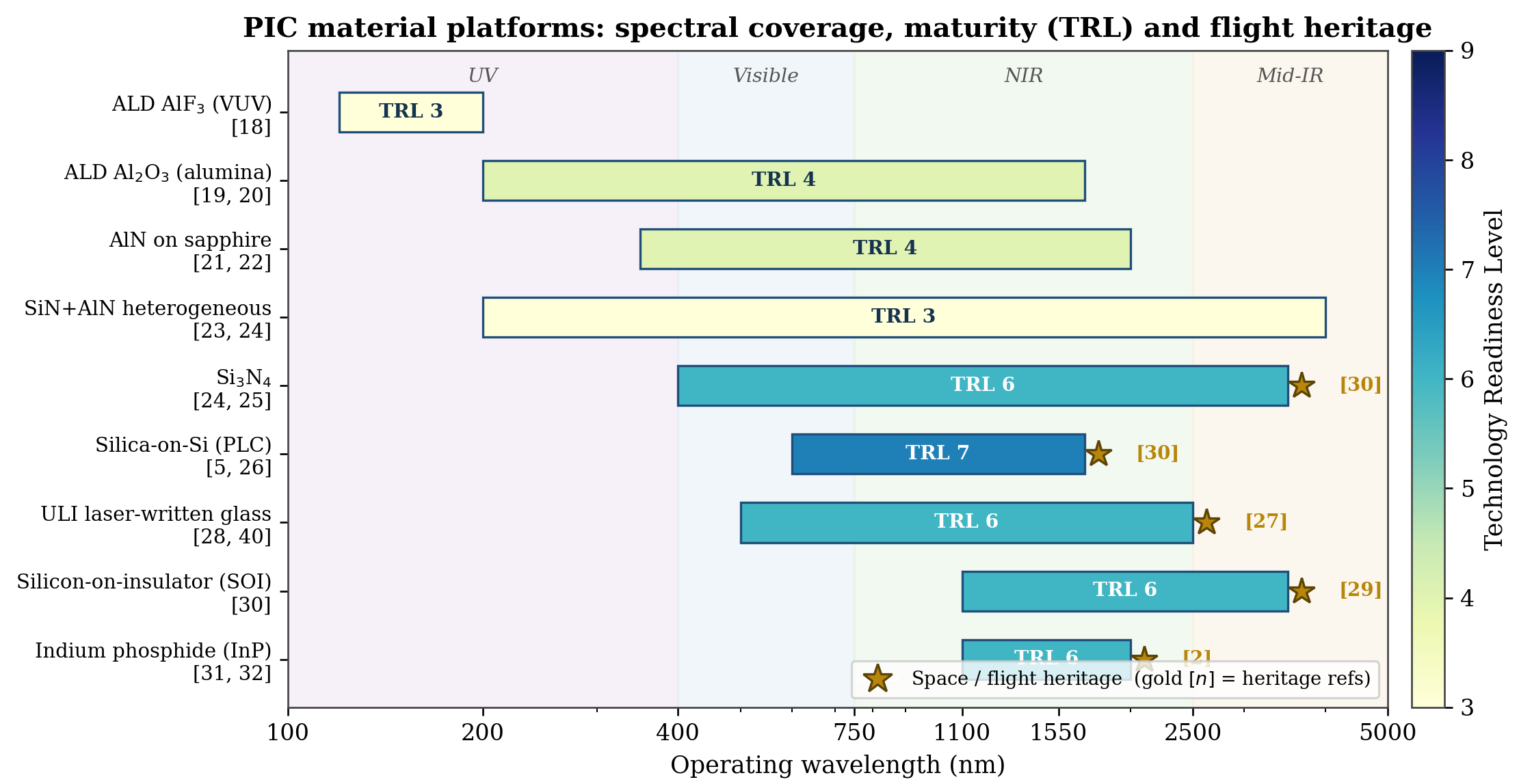}
\caption{Spectral coverage, technology-readiness level (TRL) and demonstrated space heritage of the principal astrophotonic PIC material platforms surveyed in \S\ref{sec:platforms}. Bar colour encodes the representative TRL and stars mark platforms with flight heritage; short author--year tags beside each platform indicate the supporting references. Values are representative, synthesised from the platform survey and readiness outlook \cite{pilvi2016,west2019,bradley2020,soltani2018,liu2025aln,he2025,buzaverov2024,ye2023,gatkine2022,stoll2021a,piacentini2020,nayak2021wht,mao2024,yin2021,zhao2018,zhao2019,terrasanta2025}.}
\label{fig:wavelength}
\end{figure}

Table~\ref{tab:platforms} summarises the dominant PIC waveguide platforms, ordered by operating wavelength. The two natural groupings — UV-transparent and visible / NIR — are highlighted by background colour and surveyed in detail in \S\ref{sec:uvpic} and \S\ref{sec:visnirpic} respectively. Specific contributions to a given platform (silica AWGs, Si\textsubscript{3}N\textsubscript{4} microring combs, Si\textsubscript{3}N\textsubscript{4}-on-insulator AWGs, and ULI discrete beam combiners), by Astrophotonics/AIP, Potsdam, are flagged in italics in the right-hand column. Radiation-effects evidence is consolidated separately in \S\ref{sec:radsummary}. Figure~\ref{fig:wavelength} presents this landscape graphically, mapping each platform's spectral reach against its maturity and flight heritage.

\begin{landscape}
\centering\scriptsize
\setlength{\LTpre}{0pt}\setlength{\LTpost}{0pt}
\begin{longtable}{@{}p{3.0cm} p{2.5cm} p{2.0cm} p{6.0cm} p{4.5cm} p{4.5cm}@{}}
\caption{Principal PIC waveguide platforms surveyed in this review.}\label{tab:platforms}\\
\rowcolor{esablue}\textcolor{white}{\textbf{Platform}} &
\textcolor{white}{\textbf{Core / cladding}} &
\textcolor{white}{\textbf{Range (nm)}} &
\textcolor{white}{\textbf{Key parameters}} &
\textcolor{white}{\textbf{Space-flight relevance}} &
\textcolor{white}{\textbf{Representative references}}\\
\endfirsthead
\rowcolor{esablue}\textcolor{white}{\textbf{Platform}} &
\textcolor{white}{\textbf{Core / cladding}} &
\textcolor{white}{\textbf{Range (nm)}} &
\textcolor{white}{\textbf{Key parameters}} &
\textcolor{white}{\textbf{Space-flight relevance}} &
\textcolor{white}{\textbf{Representative references}}\\
\endhead

\multicolumn{6}{@{}l}{\cellcolor{section}\textbf{\textcolor{esablue}{UV-transparent platforms}}}\\

\rowcolor{uvbox} ALD aluminum oxide (alumina) & a-Al\textsubscript{2}O\textsubscript{3} / SiO\textsubscript{2} & 200--1700 & Bandgap 5.1--7.6\,eV; $n \approx 1.65$--1.72 (vis--NUV); $<3$\,dB/cm at 371\,nm; ring $Q_i > 4.7\times10^{5}$ at 405\,nm; CMOS-compatible ALD growth on 200\,mm SiO\textsubscript{2}/Si. & Enabling platform for UV-band PICs for HWO and next-generation UV spectrographs, atomic-clock physics, biosensing. No flight heritage yet. & West et al.~\cite{west2019}; Estrada-Bayona et al.~\cite{estradabayona2022}; Bradley \& Pollnau~\cite{bradley2020}.\\

\rowcolor{uvbox} ALD aluminum fluoride & a-AlF\textsubscript{3} / SiO\textsubscript{2} & $<200$ down to VUV & ALD at 100--200\,\textcelsius; $<0.2\%$ loss in 93\,nm films from visible to 200\,nm; refractive index lower than SiO\textsubscript{2} (low-index cladding role). & Candidate cladding / anti-reflection layer for VUV PICs; of interest for HWO UV bandpass. Photonic devices still at materials-characterisation level. & Pilvi et al.~\cite{pilvi2016}.\\

\rowcolor{uvbox} AlN on sapphire & c-AlN / Al\textsubscript{2}O\textsubscript{3} & 350--$\sim$2000 & Ring $Q_i > 1.7\times10^{5}$ at 638\,nm and $>2\times10^{4}$ at 369\,nm; piezoelectric and second-order nonlinear; CMOS-compatible. & Visible / NUV PIC platform; UV astrocomb calibration; low-loss AlN at 852\,nm via ALD passivation. & Soltani et al.~\cite{soltani2018}; Liu et al.~\cite{liu2025aln}; Blumenthal et al.~\cite{blumenthal2020}.\\

\rowcolor{uvbox} Sapphire-supported heterogeneous & SiN + AlN on Al\textsubscript{2}O\textsubscript{3} & UV to IR (multi-octave) & 3D heterogeneous integration of SiN and AlN on sapphire; broadband transparency; second-order nonlinearity in AlN microcavities. & Multi-octave PICs targeting UV/vis/IR sensing and inter-PIC linking; relevant to combined-band space instruments. & He et al.~\cite{he2025}; Buzaverov et al.~\cite{buzaverov2024}.\\

\multicolumn{6}{@{}l}{\cellcolor{section}\textbf{\textcolor{esablue}{Visible and near-infrared platforms}}}\\

\rowcolor{visbox} Silicon nitride (Si\textsubscript{3}N\textsubscript{4}) & SiN / SiO\textsubscript{2} & 400--3500 & Propagation loss 0.1--1\,dB/cm; ultra-high $Q$ resonators; CMOS-compatible LPCVD; widely available foundry process. Strong $\chi^{(3)}$ nonlinearity. & Visible-NIR astrophotonic spectrographs (AWGs), coherent LiDAR engines, microcomb sources for spectrograph calibration. Excellent radiation tolerance for passive devices. & Buzaverov et al.~\cite{buzaverov2024}; Gardes et al.~\cite{gardes2022}; Ye et al.~\cite{ye2023}; Lihachev et al.~\cite{lihachev2024}. \emph{Astrophotonics/AIP} Madhav et al.~\cite{madhav2024} (POCO microring frequency comb on Si\textsubscript{3}N\textsubscript{4}).\\

\rowcolor{visbox} Silica-on-silicon (PLC, doped-SiO\textsubscript{2}) & doped SiO\textsubscript{2} / SiO\textsubscript{2} & 600--1700 & Loss $<1$\,dB/cm; mature telecom AWG / coupler / splitter foundries; low birefringence; large mode size matched to single-mode fibre. & Workhorse for AWG spectrometers in astrophotonics; inherits telecom WDM reliability heritage for inter-satellite links. Astrophotonics/AIP PAWS is the first cryogenic-detector-integrated NIR astrophotonic spectrograph on this platform. & Gatkine et al.~\cite{gatkine2017}; Gatkine et al.~\cite{gatkine2022}. \emph{Astrophotonics/AIP:} Stoll et al.~\cite{stoll2020}; Stoll et al.~\cite{stoll2021a}; Stoll et al.~\cite{stoll2021b}; Madhav et al.~\cite{madhav2024}; Hernandez et al.~\cite{hernandez2024paws}.\\

\rowcolor{visbox} Si\textsubscript{3}N\textsubscript{4} core on silica-on-silicon AWG & thin SiN core in doped SiO\textsubscript{2} & 1200--1700 & Resolving power R$\sim$1300 over H-band; fibre-to-fibre throughput $\sim$23\%; crosstalk $-17$\,dB; on-chip footprint $\sim$cm-class. & First in-house astrophotonic AWGs covering full H-band; precursor for space spectrograph chips. & Gatkine et al.~\cite{gatkine2017}. \emph{Astrophotonics/AIP:} Fernando et al.~\cite{fernando2012} --- early planar Si\textsubscript{3}N\textsubscript{4}-on-insulator astrophotonic AWG design.\\

\rowcolor{nirbox} ULI laser-written glass (3D) & fs-laser modified silica / phosphate / borosilicate glass & 500--2500 & Loss 0.1--0.3\,dB/cm; 3D routing; intrinsic single-mode operation; arbitrary geometry. Used for photonic lanterns, beam combiners, 3D fan-outs. & On-sky-demonstrated nulling interferometer (GLINT / SCExAO); on-sky-demonstrated discrete beam combiner (AIP / WHT); space-qualified to LEO TID + proton dose. & Piacentini et al.~\cite{piacentini2020}; Norris et al.~\cite{norris2020glint}; Martinod et al.~\cite{martinod2021glint}. \emph{Astrophotonics/AIP:} Pedretti et al.~\cite{pedretti2018}~\cite{pedretti2018}; Nayak et al.~\cite{nayak2021wht}; Dinkelaker et al.~\cite{dinkelaker2023dbc}; Dinkelaker~\cite{dinkelaker2024}.\\

\rowcolor{nirbox} Silicon-on-insulator (SOI) & Si / SiO\textsubscript{2} (BOX) & 1100--3500 & High index contrast ($\Delta n \approx 2$); sub-\textmu m bend radii; mature foundry; integrated Ge photodiodes and modulators; $\chi^{(3)}$ nonlinear. & Telecom-band space transceivers; first comprehensive space qualification of SiPh modulators and circuits in 2024. & Mao et al.~\cite{mao2024}; Yin et al.~\cite{yin2021}.\\

\rowcolor{nirbox} Indium phosphide (InP) & InP / InGaAsP layers & 1100--2000 & Native lasers, SOAs, EAMs, photodiodes monolithic on one chip; commercial open-access foundries; output powers $> 100$\,mW. & Space-relevant for optical satellite link transmitter / receiver PICs. Active III-V parts require dedicated radiation hardness assurance. & Zhao et al.~\cite{zhao2018,zhao2019}; Terrasanta et al.~\cite{terrasanta2025}.\\

\end{longtable}
\end{landscape}

\subsection{UV-transparent waveguide platforms}\label{sec:uvpic}
The drive towards UV PICs in space hardware comes from two complementary directions: (i) high-precision astronomical spectroscopy needs UV-band wavelength calibration for HARPS3 / ESPRESSO and the future HWO UV channel; and (ii) planetary and heliospheric science needs ruggedised UV instrumentation. Silicon (E\textsubscript{g} = 1.1\,eV) is unusable below the near-infrared; silicon nitride extends usefully into the visible but begins to fail at $\sim$400\,nm. Two materials families bridge the gap into the UV: wide-bandgap nitrides (AlN, GaN) and amorphous oxides / fluorides grown by atomic layer deposition~\cite{blumenthal2020}.

\subsubsection{ALD-grown aluminum oxide (amorphous alumina)}
The most mature low-loss UV PIC platform reported in the open literature is fully etched aluminum-oxide waveguides grown by ALD on thermal silicon dioxide. West et al.~\cite{west2019} reported the first detailed study, showing single-transverse-mode operation with $<3$\,dB/cm propagation loss at 371\,nm and ring-resonator intrinsic quality factors exceeding $4.7\times10^{5}$ at 405\,nm. The thermo-optic coefficient was measured as $2.75\times10^{-5}$\,RIU/\textcelsius. Amorphous Al\textsubscript{2}O\textsubscript{3} has an electronic bandgap between 5.1 and 7.6\,eV depending on the deposition mechanism, corresponding to transparency down to 163--243\,nm; transmission losses below 4\,dB/cm at 250\,nm have been reported on ALD films deposited on fused silica. The refractive index of 1.65--1.72 in the visible / NUV is conveniently higher than that of SiO\textsubscript{2}, enabling standard buried-waveguide architectures with thermal-oxide cladding.

Estrada-Bayona et al.~\cite{estradabayona2022} extended this to functional UV PICs for far-field structured illumination autofluorescence microscopy, demonstrating 700\,nm-wide single-mode waveguides etched in 120\,nm of ALD AlO\textsubscript{x} with 3\,dB/cm loss at 360\,nm using an air top-cladding. Compatibility with 200 / 300\,mm CMOS-grade thermal-oxide wafers makes the platform attractive for foundry adoption. Rare-earth-doped Al\textsubscript{2}O\textsubscript{3} provides on-chip gain and lasing on the same platform with continuous transparency from 150 to 5500\,nm~\cite{bradley2020}.

For space, the dominant outstanding development is a flight-grade hermetic package and a documented radiation hardness assurance (RHA) campaign on ALD alumina — neither is yet in the open literature. The 2023 Astrophotonics Roadmap explicitly identifies UV-extended platforms as one of its priority technology-development items~\cite{jovanovic2023}.

\subsubsection{ALD-grown aluminum fluoride (AlF\textsubscript{3})}
Aluminum fluoride extends the transparency window further into the vacuum-UV. Pilvi et al.~\cite{pilvi2016} developed an ALD process for AlF\textsubscript{3} using trimethylaluminum and anhydrous HF at substrate temperatures 100--200\,\textcelsius, achieving amorphous films with characterised optical losses below 0.2\% (93\,nm thickness) from visible wavelengths down to 200\,nm. X-ray photoelectron spectroscopy showed no residual aluminum-oxide signature. The refractive index of AlF\textsubscript{3} is significantly lower than that of SiO\textsubscript{2}, so the immediate device-level role is as an anti-reflection coating or as a low-index cladding for higher-index UV cores. Full PIC-level demonstrations beyond materials characterisation are not yet in the peer-reviewed literature; the material remains one of the strongest candidate claddings for HWO-class VUV instrumentation.

\subsubsection{Aluminum nitride on sapphire and sapphire-supported platforms}
Crystalline AlN grown on c-cut sapphire has a 6.2\,eV bulk bandgap and is transparent from $\sim$210\,nm well into the IR. Soltani et al.~\cite{soltani2018} demonstrated nanocrystalline AlN-on-sapphire waveguides with ring resonator intrinsic $Q$ exceeding $1.7\times10^{5}$ at 638\,nm and $>2\times10^{4}$ at 369.5\,nm. AlN also offers electro-optic modulation at multi-GHz frequencies and intrinsic second-order nonlinearity for on-chip frequency conversion (Blumenthal et al.~\cite{blumenthal2020}). Liu et al.~\cite{liu2025aln} reported a record-low 2\,dB/cm at 852\,nm for AlN-on-sapphire waveguides using ALD-Al\textsubscript{2}O\textsubscript{3} passivation and post-cladding rapid thermal annealing.

A more recent direction is 3D heterogeneous integration on sapphire: He et al.~\cite{he2025} vertically integrated Si\textsubscript{3}N\textsubscript{4} and AlN PIC layers on a sapphire substrate, exploiting AlN for nonlinear conversion and Si\textsubscript{3}N\textsubscript{4} for low-loss passive routing within a single multi-octave (UV to IR) platform. Sapphire-supported architectures inherit the optical and thermal robustness of bulk sapphire — an attractive property for ruggedised space instruments. A comprehensive UV-to-IR integration survey across nitride and oxide platforms is given by Blumenthal et al.~\cite{blumenthal2020}.

\subsubsection{Lithium niobate for chip-scale UV generation}
A complementary route to chip-scale UV light is non-linear conversion in periodically-poled, nano-fabricated thin-film lithium niobate (LiNbO\textsubscript{3}, LN) waveguides. Ludwig et al.~\cite{ludwig2024} demonstrated a UV astronomical-spectrograph calibration source by combining an LN second- and sum-frequency generator with a microresonator electro-optic comb, providing a route to UV calibration for next-generation ground- and space-based precision spectrographs.

\subsection{Visible and near-infrared waveguide platforms}\label{sec:visnirpic}

\subsubsection{Silicon nitride (Si\textsubscript{3}N\textsubscript{4})}
Si\textsubscript{3}N\textsubscript{4} PICs are the dominant low-loss visible / NIR integrated photonics platform. Comprehensive reviews are provided by Buzaverov et al.~\cite{buzaverov2024} (visible to mid-IR, foundry-grade processes) and Gardes et al.~\cite{gardes2022} (hybrid integration). Propagation losses below 1\,dB/cm at 1550\,nm are routine, and ultra-low-loss processes reach below 1\,dB/m for $\sim$\textmu m-thick waveguides (Ye et al.~\cite{ye2023}). CMOS-compatible LPCVD on 200--300\,mm wafers gives access to commercial foundries, with hybrid and heterogeneous integration of III-V active devices (lasers, SOAs) demonstrated for soliton microcombs and ultra-low-noise lasers (Lihachev et al.~\cite{lihachev2024}).

For space applications the principal attractions are (i) low loss extending down to $\sim$400\,nm (vs.\ $\sim$1100\,nm for SOI); (ii) the same chip can host both passive optical processing and on-chip microcomb sources for high-precision spectrograph calibration; and (iii) excellent radiation tolerance of passive components inherited from telecom-grade reliability data (Yin et al.~\cite{yin2021}). A typical qualification flow for a Si\textsubscript{3}N\textsubscript{4} PIC in the present document maps row-by-row onto the master table of \S\ref{sec:tailoring}, with the radiation-test budget effectively relaxed for the passive core (see \S\ref{sec:radsummary}).

\paragraph{Astrophotonic AWGs in Si\textsubscript{3}N\textsubscript{4}.} Gatkine et al.~\cite{gatkine2017} fabricated H-band ($\lambda$ = 1450--1650\,nm) AWG spectrometers on a silica-on-silicon substrate using a thin Si\textsubscript{3}N\textsubscript{4} core. These devices achieved a peak fibre-to-fibre throughput of 23\%, resolving power $R \approx 1300$, free spectral range 10\,nm and channel crosstalk of $-17$\,dB. Subsequent broadband AWGs spanning 1200--1650\,nm in both doped-SiO\textsubscript{2} and Si\textsubscript{3}N\textsubscript{4} platforms have been compared by Gatkine et al.~\cite{gatkine2022}, targeting low-resolution astrophotonic back-end spectrographs for photonic lanterns and nulling interferometers.

\paragraph{Si\textsubscript{3}N\textsubscript{4} for coherent LiDAR engines.} Lihachev et al.~\cite{lihachev2024} demonstrated a photonic-electronic coherent LiDAR engine combining a hybrid Si\textsubscript{3}N\textsubscript{4} circuit, a tunable Vernier laser with piezoelectric actuators, an erbium-doped waveguide amplifier and a co-integrated 130\,nm SiGe BiCMOS arbitrary-waveform generator — a strong demonstration of full-stack PIC LiDAR with foundry compatibility, directly relevant to spacecraft proximity-operations and rendezvous instruments.

\paragraph{Si\textsubscript{3}N\textsubscript{4} microresonator astrocombs (Astrophotonics/AIP Potsdam).} The Astrophotonics/AIP group has developed an integrated Si\textsubscript{3}N\textsubscript{4} micro-ring-resonator-based astronomical optical frequency comb generator. Bodenmuller et. al, \cite{Bodenmueller2025FrequencyCombs, Bodenmueller2020OFCMicroRing, Bodenmueller2023POCO} reported fabrication and spectral characterisation of Si\textsubscript{3}N\textsubscript{4} ring resonators with 250 and 500\,\textmu m diameter, and 28.55-GHz repetition-rate combs in a Si\textsubscript{3}N\textsubscript{4} microring via amplitude-modulated pumping. The mature instrument-level POCO comb integrated with PAWS via the AIP campus fibre network was reported by Madhav et al.~\cite{madhav2024}, demonstrating that a microring frequency comb can serve as a remote calibration source for an astrophotonic spectrograph over a 22\,km fibre link. 

\subsubsection{Silica-on-silicon planar lightwave circuits (PLC)}
Silica-on-silicon (PLC) technology — doped-SiO\textsubscript{2} cores in undoped-SiO\textsubscript{2} cladding on a silicon carrier — was the original commercial PIC platform and remains the lowest-loss available. Gatkine et al.~\cite{gatkine2022} reported doped-SiO\textsubscript{2} AWGs spanning 1200--1650\,nm with fibre-to-fibre throughput up to 79\% (1\,dB) and resolving power $R \approx 200$ for ultra-broadband astrophotonic applications. PLC components have benefited from decades of fibre-optic-telecom reliability work (Telcordia GR-1221, GR-468) that is directly applicable to space tailoring (see also \S\ref{sec:fibers}).

A substantial body of dedicated astrophotonic AWG development on the silica-on-silicon platform has been carried out at the Leibniz-Institut für Astrophysik Potsdam (AIP). Stoll et al.~\cite{stoll2020} examined the impact of phase errors on high-resolution AWG performance and phase-error trimming on a custom silica AWG. Stoll et al.~\cite{stoll2021a} presented the first-generation custom silica AWGs in the H-band with resolving powers up to $R \approx 18\,900$, and Stoll et al.~\cite{stoll2021b} extended the design to low-aberration three-stigmatic-point layouts. The first instrument-level integration --- the \emph{Potsdam Arrayed Waveguide Spectrograph} (PAWS), a silica AWG coupled to a Hawaii2RG detector cooled to $-190$\,\textcelsius\ --- was reported by Madhav et al.~\cite{madhav2024}, alongside the companion \emph{Potsdam Comb} (POCO) microresonator frequency comb on Si\textsubscript{3}N\textsubscript{4} used to calibrate PAWS through a 22\,km fibre link. Preliminary on-sky-relevant characterisation of PAWS was reported by Hernandez et al.~\cite{hernandez2024paws}. Earlier conceptual work on AWGs and silicon-nitride-on-insulator alternatives appears in Fernando et al.~\cite{fernando2012}.

\subsubsection{Ultrafast laser-inscribed (ULI) glass waveguides}
Femtosecond laser direct-writing of single-mode waveguides in bulk silica or phosphate glass — ultrafast laser inscription, ULI — enables genuinely three-dimensional photonic circuits that are difficult or impossible to fabricate by planar lithography. Losses of 0.1--0.3\,dB/cm are routine. The ULI platform underpins photonic lanterns, 3D beam combiners and the GLINT family of nulling interferometers (Norris et al.\ and Martinod et al.~\cite{norris2020glint,martinod2021glint}). Laser-written tri-couplers and 3D pupil remappers have flown on-sky at the Subaru Telescope.

A particularly important contribution to the space readiness of ULI is the work of Piacentini et al.~\cite{piacentini2020}, who exposed ULI straight waveguides, directional couplers and Mach--Zehnder interferometers in glass to proton and gamma-ray doses representative of low-Earth orbit and observed no measurable change in their performance — a foundational step towards qualifying ULI devices for satellite-based quantum-communication and astrophotonic missions. This study is one of only two open-literature space-qualification campaigns referenced explicitly across the qualification matrix of \S\ref{sec:tailoring}, rows 13--15.

ULI in borosilicate glass underpins the discrete-beam-combiner (DBC) effort at AIP. Pedretti et al.~\cite{pedretti2018} reported the J-band six-input DBC concept. Dinkelaker et al.~\cite{dinkelaker2023dbc} reported the first six-telescope DBC for J-band stellar interferometry with $\sim 56\%$ throughput, the highest combiner count among existing DBCs. Nayak et al.~\cite{nayak2021wht} demonstrated first stellar photons for an integrated-optics DBC at the William Herschel Telescope. A broader review of astrophotonic ULI activity, with explicit emphasis on the European pipeline towards space-qualified components, is given by Dinkelaker 2024.

\subsubsection{Silicon-on-insulator (SOI) silicon photonics}
SOI silicon photonics underpins the bulk of the terrestrial data-centre transceiver market, and has recently been the subject of a dedicated space-qualification campaign. Mao et al.~\cite{mao2024} demonstrated that silicon photonic modulators and Ge-on-Si photodiodes manufactured in a commercial foundry can be characterised against the full SEE / TID / DDD test matrix of avionics-grade flight components, with acceptable degradation up to mission-relevant doses for LEO and beyond. Yin et al.~\cite{yin2021} provide a detailed account of high-energy proton, neutron and Co-60 gamma-ray effects on passive silicon photonic devices, with refractive-index shifts of microring and Mach--Zehnder structures being the dominant observable. The recent review by Terrasanta et al.~\cite{terrasanta2025} consolidates these results for the optical-satellite-link community.

The general picture is that passive SOI structures are radiation tolerant to mission-end LEO/GEO doses with modest design margin; the active components (Ge-on-Si photodiodes, p-i-n modulators, supporting electronic drivers) carry the radiation-hardness burden.

\subsubsection{Indium phosphide (InP)}
InP is the only monolithic platform that natively integrates laser sources, SOAs, modulators and photodiodes on a single chip. Open-access commercial foundries now offer multi-project-wafer access. Zhao et al.~\cite{zhao2018} demonstrated InP transmitter PICs for free-space optical communications, integrating a sampled-grating DBR laser with a high-speed SOA, a Mach-Zehnder modulator and a booster SOA, achieving 7\,Gbps error-free operation. Zhao et al.~\cite{zhao2019} extended this to high-power transmitter platforms producing $> 230$\,mW off-chip. For optical satellite links, Terrasanta et al.~\cite{terrasanta2025} reviewed InP PIC architectures and the corresponding radiation-tolerance literature; displacement damage in active III-V components is the principal concern, which maps directly onto the DDD/proton row of the master qualification table.

\subsection{Application case studies}\label{sec:app_cases}

\begin{figure}[htbp]
\centering
\includegraphics[width=\linewidth]{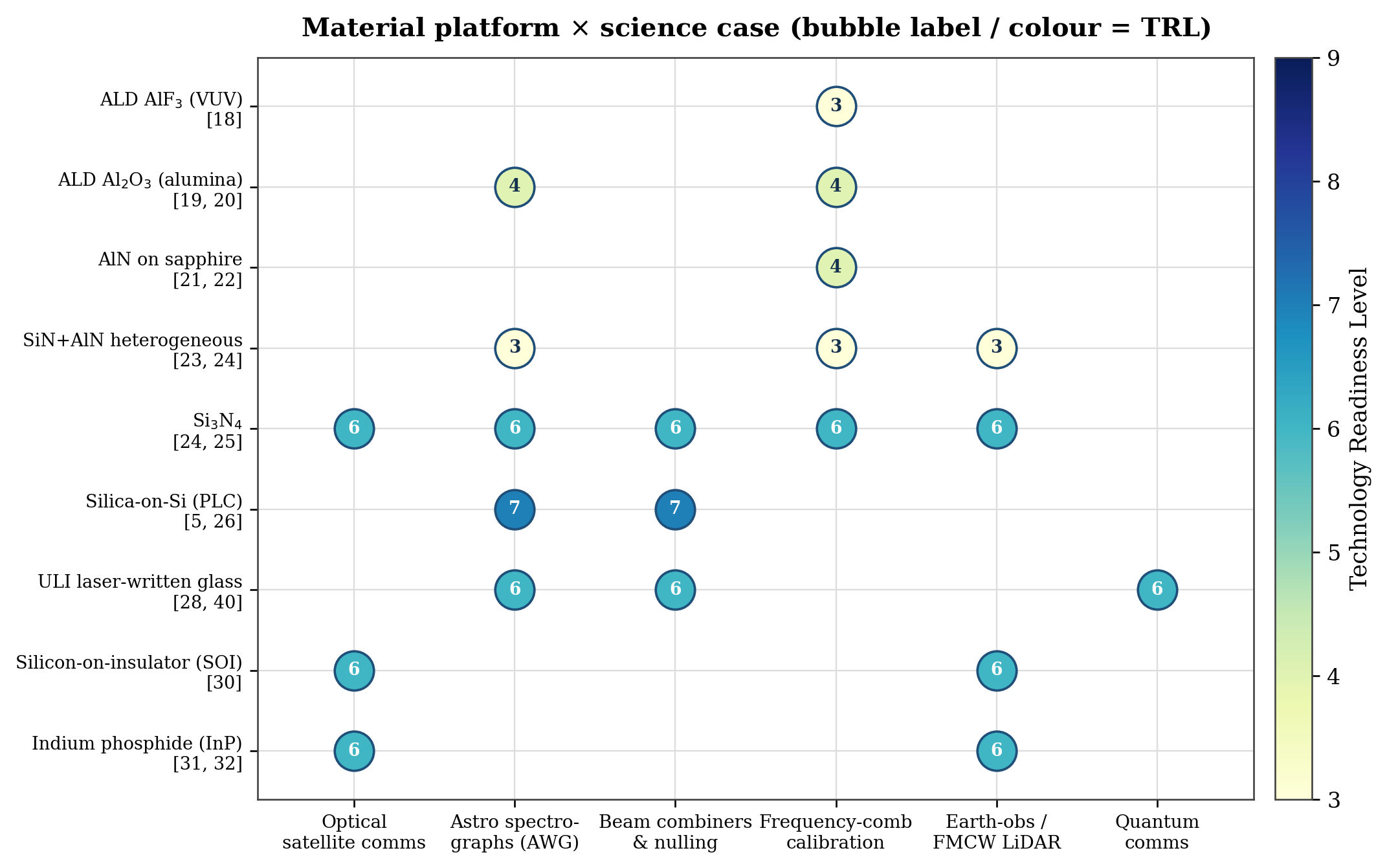}
\caption{Mapping of PIC material platforms onto the principal astrophotonic and space-photonic cases. 
.}
\label{fig:sciencecase}
\end{figure}

Figure~\ref{fig:sciencecase} summarises which material platforms address each of the principal science cases discussed below, colour-coded by technology-readiness level.

\subsubsection{Optical satellite communications}
NASA's Laser Communications Relay Demonstration (LCRD), in GEO since 2021, demonstrates EDFA-amplified 1064 / 1550\,nm bidirectional links between ground and a GEO node. LCRD uses fibre-coupled amplifier modules rather than fully integrated PIC transceivers, but provides the qualified-component infrastructure (high-power laser-diode arrays, EDFAs, photodetectors) on which future PIC-based architectures will build.

TBIRD — the TeraByte Infrared Delivery 6U CubeSat launched in 2022 — hosts two commercial off-the-shelf 100\,Gbps coherent fibre-optic transceivers used in wavelength-division multiplexing, achieving error-free transfers of more than 1\,TB per 5-minute pass to ground (Schieler et al.~\cite{schieler2023}). TBIRD demonstrates that mature terrestrial-foundry-derived coherent transceivers — themselves PIC-based at chip level — survive LEO and deliver flagship-class data rates from a CubeSat platform.

NASA's Deep Space Optical Communications (DSOC) demonstration, launched on the Psyche mission in October 2023, achieved laser communications at 218 million miles by 2025 (Biswas et al.~\cite{biswas2024dsoc}). The flight laser transceiver uses a 1064\,nm uplink receive channel and a 1550\,nm serially-concatenated pulse-position-modulated downlink; the ground receiver is a superconducting nanowire single-photon detector array. The flight side of DSOC is not a single-chip PIC but draws heavily on commercial PIC-foundry components for laser-amplifier and detector modules.

The most comprehensive synthesis of PIC technology status for satellite communications, including side-by-side comparison of InP, SOI and Si\textsubscript{3}N\textsubscript{4} architectures and their respective space-effects experimental records, is given by Terrasanta et al.~\cite{terrasanta2025}.

\subsubsection{Astrophotonic instrumentation}
The 2023 Astrophotonics Roadmap (Jovanovic et al.~\cite{jovanovic2023}) is the community consensus document for PICs in astronomical instrumentation. It identifies 24 development areas across design tools, fabrication processes, integration, hybridisation and component-level characterisation, and explicitly calls out the need for radiation-tested and space-qualified astrophotonic devices. A broader review of astrophotonic technologies — photonic lanterns, fibre Bragg gratings, AWGs, beam combiners, frequency combs — is given by Madhav et al.~\cite{roth2023spie}, which surveys the AIP pipeline from concept through to instrument-level integration of PAWS and POCO. A 2024 update of the AIP astrophotonic instruments programme is given by Madhav et al.~\cite{madhav2024}.

\begin{itemize}[leftmargin=1.4em,nosep]
  \item \textbf{Photonic lanterns and integral-field units.} The photonic lantern (Leon-Saval et al.~\cite{leonsaval2010}) adiabatically converts between a multimode input and an array of single-mode outputs, enabling diffraction-limited spectroscopy with multimode-fibre input efficiency. It has become a foundational building block; combinations with FBG OH-line filters, AWGs and ULI 3D fan-outs are surveyed in Jovanovic et al.~\cite{jovanovic2023} and Madhav et al.~\cite{roth2023spie}. The Astrophotonics/AIP group has investigated optimal SMF packing arrangements for photonic lanterns (Davenport et al.~\cite{davenport2021}) and mode-expansion theory in step-index multimode fibres for astronomical spectroscopy (Hernandez et al.~\cite{hernandez2021mode}).
  \item \textbf{Arrayed-waveguide-grating spectrographs.} H-band AWGs on silica-on-silicon with Si\textsubscript{3}N\textsubscript{4} core (Gatkine et al.~\cite{gatkine2017}) and broadband 1200--1650\,nm AWGs in both doped-SiO\textsubscript{2} and Si\textsubscript{3}N\textsubscript{4} (Gatkine et al.~\cite{gatkine2022}) are the closest existing antecedents of a space-spectrograph chip. The Astrophotonics/AIP PAWS instrument (Madhav et al.~\cite{madhav2024}) demonstrates an end-to-end spectrograph on the silica platform, integrated with a cryogenically cooled Hawaii2RG detector and a remote POCO frequency-comb calibration source. Higher-resolution (R$>10\,000$) AWGs and full-band stitched FSR remain explicit priorities in the 2023 Astrophotonics Roadmap, building directly on the Stoll et al.~\cite{stoll2020,stoll2021a,stoll2021b} Gen-I and Gen-II silica AWG demonstrations.
  \item \textbf{Nulling interferometers and discrete beam combiners.} Norris et al.~\cite{norris2020glint} reported the first on-sky demonstration of an integrated-photonic nulling interferometer at the Subaru Telescope, achieving null-depth precision of $\sim 10^{-4}$ on sky and stellar angular diameters to milliarcsecond accuracy. Martinod et al.~\cite{martinod2021glint} extended this to a dispersed multi-baseline architecture with null depths better than $10^{-3}$ in laboratory conditions. In parallel, the Astrophotonics/AIP group (Pedretti, Dinkelaker, Nayak, Madhav, Roth and collaborators) has developed ULI-based Discrete Beam Combiner (DBC) chips, with the first six-telescope DBC for J-band interferometry demonstrated by Dinkelaker et al.~\cite{dinkelaker2023dbc} and first stellar photons at the WHT reported by Nayak et al.~\cite{nayak2021wht}. The PIC nuller and DBC architectures are directly applicable to future flagship space missions (HWO, LIFE).
  \item \textbf{Frequency-comb spectrograph calibration.} Ludwig et al.~\cite{ludwig2024} demonstrated UV astrocomb generation by chip-integrated nonlinear photonics in periodically-poled nanofabricated LN waveguides combined with electro-optic and microresonator combs, providing a path to chip-scale UV calibration for HARPS3, ESPRESSO and future space precision spectrographs. The Si\textsubscript{3}N\textsubscript{4} microring POCO comb developed at Astrophotonics/AIP~\cite{madhav2024} provides the complementary NIR / visible-band calibration capability, demonstrated over a 22\,km campus fibre network at AIP.
\end{itemize}

\subsubsection{Earth-observation lidar and FMCW coherent LiDAR}
Although the ALADIN (Aeolus, 2018) and ATLID (EarthCARE, 2024) atmospheric-lidar transmitters use diode-pumped Nd:YAG solid-state lasers rather than full PICs, the high-power laser-diode pump arrays are themselves space-qualified semiconductor laser components. The natural next step is PIC-based FMCW coherent LiDAR: Martin et al.~\cite{martin2018} demonstrated a silicon-platform PIC FMCW LiDAR with on-chip waveform calibration, balanced detectors and beam steering, ranging out to 60\,m with $<5$\,mW output. Lihachev et al.~\cite{lihachev2024} extended this to a foundry-compatible photonic-electronic LiDAR engine on Si\textsubscript{3}N\textsubscript{4}.

\subsubsection{Quantum communications and CubeSat experiments}
The space qualification of ULI integrated waveguides by Piacentini et al.~\cite{piacentini2020} was motivated by satellite-based quantum-communication payloads. Laser-written interferometers, directional couplers and Mach-Zehnder devices in glass were shown to survive LEO-equivalent proton and gamma-ray exposure without measurable change in their characteristics — establishing a route towards satellite QKD payloads built from monolithic ULI circuits rather than discrete fibre components.

\subsection{Radiation-effects summary by platform}\label{sec:radsummary}
Table~\ref{tab:radsummary} consolidates the open-literature radiation-effects data for the platforms surveyed. Three points emerge consistently:

\begin{enumerate}[leftmargin=1.6em,nosep]
  \item \textbf{Passive low-index-contrast platforms} (silica-on-silicon, ULI glass, Si\textsubscript{3}N\textsubscript{4}) are essentially radiation-immune at typical LEO TID and proton fluences (Piacentini et al.~\cite{piacentini2020}; Yin et al.~\cite{yin2021}).
  \item \textbf{Passive SOI structures} display small but measurable wavelength-shifts of microring resonators due to refractive-index changes in Si core and SiO\textsubscript{2} cladding under high-dose gamma exposure ($\sim$15\,Mrad); below 1\,Mrad the shifts are within typical thermal-tuning capability~\cite{yin2021,mao2024}.
  \item \textbf{Active components} — Ge-on-Si photodiodes, p-i-n modulators, and III-V active devices (InP lasers, SOAs) — are the principal radiation-hardness limit. Ge-on-Si dark-current rises and III-V threshold-current drift dominate the displacement-damage budget (Arnold et al.~\cite{arnold2022}; Mao et al.~\cite{mao2024}; Terrasanta et al.~\cite{terrasanta2025}).
\end{enumerate}

\begin{table}[ht]
\centering\footnotesize
\caption{Open-literature radiation-effects summary for the PIC platforms surveyed. Cross-reference to the master qualification table of \S\ref{sec:tailoring} rows 13--15 and to the tailoring matrix of \S\ref{sec:tailoring} rows 13--15.}\label{tab:radsummary}
\begin{tabularx}{\textwidth}{@{}p{3.0cm} p{4.0cm} p{4.0cm} X@{}}
\toprule
\rowcolor{esablue}\textcolor{white}{\textbf{Platform}} & \textcolor{white}{\textbf{TID (Co-60)}} & \textcolor{white}{\textbf{Proton / DDD}} & \textcolor{white}{\textbf{SEE / single-event}}\\
\midrule
ULI glass waveguides & No change to 6\,kGy gamma (Piacentini 2020). & No change to $6\times10^{10}$\,p/cm\textsuperscript{2} at 60\,MeV (Piacentini 2020). & Not applicable (passive).\\
\rowcolor{rowalt} Silica-on-silicon AWG / PLC & Tolerant to $\geq$1\,Mrad (Yin 2021). & Tolerant to LEO-equivalent proton fluence (Yin 2021). & Not applicable (passive).\\
Si\textsubscript{3}N\textsubscript{4} passive & Small wavelength shifts at $>1$\,Mrad; recoverable (Yin 2021). & Negligible to LEO fluences (Yin 2021). & Not applicable.\\
\rowcolor{rowalt} SOI silicon photonics — passive & MRR / MZI wavelength shifts at multi-Mrad doses (Yin 2021). & Index changes scale with DDD (Yin 2021; Mao 2024). & Negligible for passive structures.\\
SOI — Ge photodiodes / modulators & Modest dark-current rises to 4\,dB at high TID (Arnold 2025). & Dominant degradation mechanism; dark current $\propto$ NIEL$\times$fluence (Arnold 2025; Mao 2024). & SET on PDs / TIAs documented (Mao 2024).\\
\rowcolor{rowalt} InP active PICs & Laser threshold drift; dependent on epitaxy and confinement factor (Terrasanta 2025). & Threshold-current rise, slope-efficiency drop with proton fluence (Terrasanta 2025). & SEU / SEFI on driver electronics; SEL screening required.\\
ALD-grown alumina / AlF\textsubscript{3} & Not yet reported in open space-qualification literature. & Not yet reported. & Not applicable (passive).\\
\rowcolor{rowalt} AlN-on-sapphire & Not yet reported in space-qualification literature. & Not yet reported. & Not applicable.\\
\bottomrule
\end{tabularx}
\end{table}

\subsection{Outlook and platform-specific qualification gaps}\label{sec:platform_outlook}

\begin{figure}[htbp]
\centering
\includegraphics[width=\linewidth]{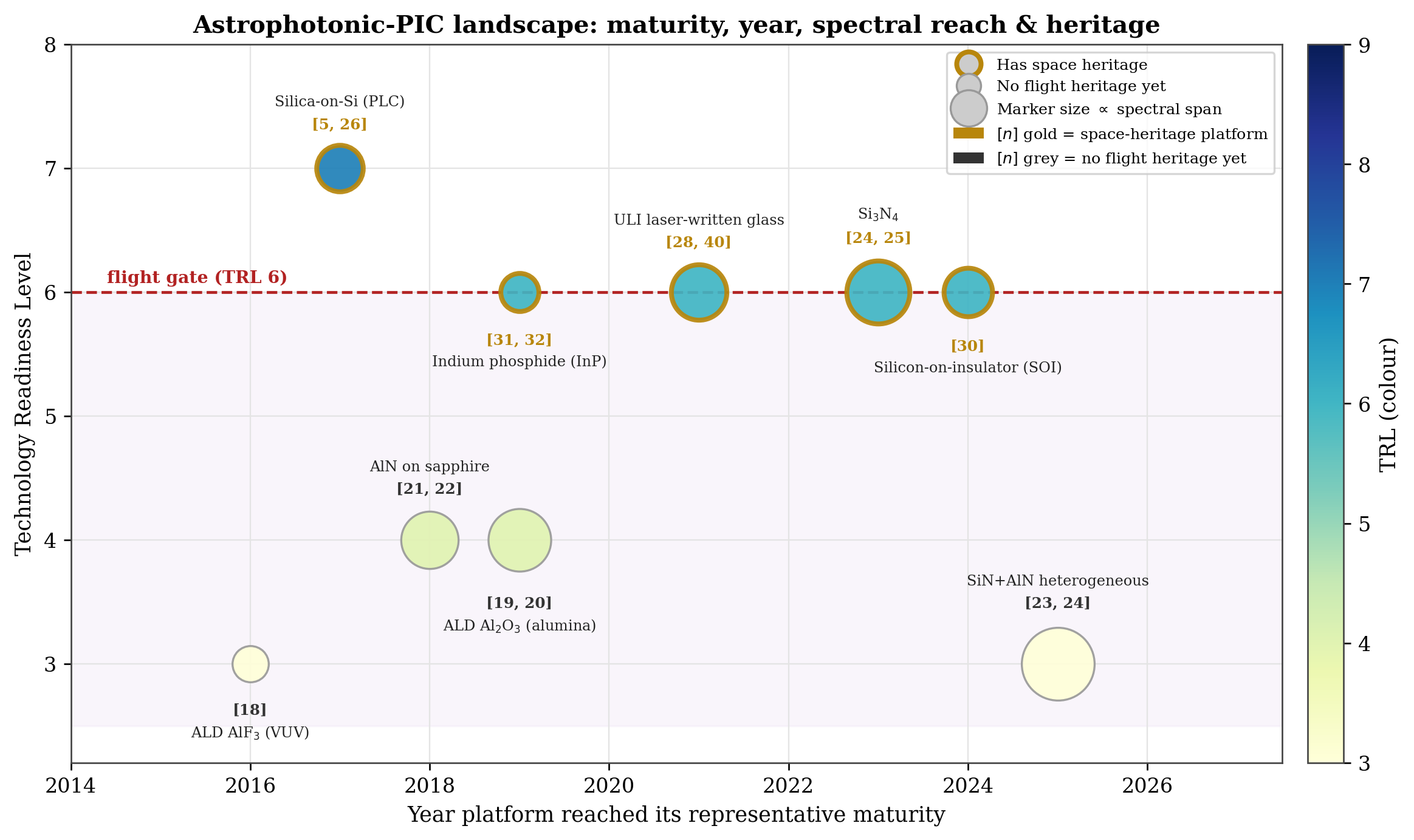}
\caption{Integrated maturity landscape of the surveyed PIC platforms. 
}
\label{fig:landscape}
\end{figure}

Figure~\ref{fig:landscape} consolidates the surveyed platforms in a single maturity-versus-year view, tying together material, wavelength, TRL, year and flight heritage. The literature reviewed above paints a clear picture: visible / NIR PIC platforms (Si\textsubscript{3}N\textsubscript{4}, silica-on-silicon, ULI glass, SOI, InP) sit at TRL 5--7 for several space applications, with the 2024 Mao et al.\ campaign on silicon photonics and the 2020/2021 Piacentini/Vogl/Corrielli campaign on ULI representing the two most concrete pieces of space-qualification evidence in the open literature. UV-transparent platforms (ALD alumina, AlF\textsubscript{3}, AlN-on-sapphire) are at TRL 3--4 with strong fundamental materials demonstrations (West et al., Pilvi et al., Soltani et al., and He et al.~\cite{west2019,pilvi2016,soltani2018,he2025}) but no published space qualification.

The principal gaps identified consistently across the open literature and the 2023 Astrophotonics Roadmap — and which therefore drive the platform-specific tailoring of the master qualification flow of \S\ref{sec:tailoring} and the TRL roadmap of \S\ref{sec:trl} — are:

\begin{itemize}[leftmargin=1.4em,nosep]
  \item \textbf{Hermetic spaceflight packaging of UV PICs} --- die-bond, fibre-attach and lid-seal processes proven for $<\!400$\,nm operation, including UV-tolerant epoxies or solder fixturing.
  \item \textbf{Radiation hardness assurance campaigns} on ALD-alumina and AlF\textsubscript{3} platforms covering proton, neutron, heavy-ion and combined-proton irradiation per ECSS-Q-ST-60-15C Rev.\,1 / NASA-STD-8739.11 (see also the practitioner notes in \S\ref{sec:practitioner}).
  \item \textbf{Cryogenic operation} of Si\textsubscript{3}N\textsubscript{4}, ULI and SOI platforms for IR astrophotonic spectrographs (HWO, Origins, LIFE) --- phase and dark-current stability at 80\,K is poorly characterised in the open literature; this drives the cryogenic-cycling row of the TRL matrix in \S\ref{sec:trl}.
  \item \textbf{Hybrid integration} of III-V active components onto wide-bandgap waveguide platforms (Si\textsubscript{3}N\textsubscript{4}, AlN-on-sapphire) with space-qualified bonding and thermal-management solutions.
  \item \textbf{High-power UV-band sources} on chip --- chip-scale frequency conversion via LN, KTP or AlN second-order processes for UV calibration lines, with output power stable over orbital cycling.
  \item \textbf{Foundry access to space-grade PDKs} --- most flight demonstrations to date have used research-grade or repurposed commercial-foundry runs without an explicit space PDK or qualification heritage.
\end{itemize}

For the Habitable Worlds Observatory and similar UV / optical / IR flagships, the most immediate-value developments are: (i) a UV-band Si\textsubscript{3}N\textsubscript{4} or alumina micro-comb / electro-optic-comb chip for spectrograph wavelength calibration; (ii) ULI glass beam-combiner chips for the nulling-interferometer architecture; (iii) packaged Si\textsubscript{3}N\textsubscript{4} AWG modules covering 0.5--1.7\,\textmu m with sub-pm phase stability; and (iv) a documented hermetic packaging and qualification flow that closes the loop with ECSS-Q-ST-60C / EEE-INST-002 part-screening practice.

\section{Practitioner notes — JPL/NASA project experience}\label{sec:practitioner}
The formal qualification flow of \S\ref{sec:tailoring} maps the \textbf{complete} set of ECSS, GSFC-STD-7000B and EEE-INST-002 verifications. In day-to-day practice on a JPL/NASA-led astrophotonic R\&D project the test campaign typically reduces to \textbf{three} principal activity blocks for an electronic or active-photonic component, with the remaining items absorbed either at part-vendor level (incoming inspection, hermeticity, burn-in) or at instrument level (EMC, contamination control):

\begin{enumerate}[leftmargin=1.6em,nosep]
  \item \textbf{Radiation testing} — almost always carried out with the device under bias (``powered'').
  \item \textbf{Mechanical shock and vibration.}
  \item \textbf{Thermal cycling.}
\end{enumerate}

For the European framework the relevant umbrella is \textbf{ECSS — European Cooperation for Space Standardization}. The ECSS-Q-ST-60 and -70 series cover the same ground as the NASA standards but with different naming conventions. The notes in this section concentrate on JPL/NASA-oriented practice as encountered in current astrophotonic technology projects; they are a working summary, not a formal cross-walk to ECSS, which remains the authoritative ESA standard for radiation hardness assurance (in particular ECSS-Q-ST-60-15C Rev.\,1, Mar 2025).

\subsection{Radiation testing — three sub-tests}

\renewcommand{\arraystretch}{1.3}
\noindent\begin{tabularx}{\textwidth}{@{}p{1.5cm} p{3.5cm} X@{}}
\toprule
\rowcolor{esablue}
\textcolor{white}{\textbf{Sub-test}} & \textcolor{white}{\textbf{What is tested}} & \textcolor{white}{\textbf{Source / particle and notes}}\\
\midrule
\textbf{1A} & Single-event effects (SEU, SET, SEFI) and \textbf{destructive} single-event latch-ups (SEL) & Heavy-ion accelerator beams up to a defined maximum linear energy transfer (LET). Maximum LET specified by the agency: \textbf{JPL: 75\,MeV$\cdot$cm\textsuperscript{2}/mg}; \textbf{ESA / ECSS: 60\,MeV$\cdot$cm\textsuperscript{2}/mg}. Different ion species deliver different LET — heavier ions and higher particle energies give higher LET. The figure that counts for pass/fail is the LET \textbf{at the device surface}, i.e.\ after the beam window and any package coverage.\\
\rowcolor{rowalt}
\textbf{1B} & Total Non-Ionizing Dose (TNID) / Displacement Damage Dose (DDD) — accumulated displacement damage in the semiconductor crystal lattice & Classically performed with \textbf{neutrons}; modern practice often uses \textbf{protons} (a closer match to the orbital environment, see below). DDD primarily affects active III-V parts (laser threshold rise, photodiode dark-current increase). Passive SiO\textsubscript{2}/Si\textsubscript{3}N\textsubscript{4}/ULI PIC waveguides are essentially immune.\\
\textbf{1C} & Total Ionizing Dose (TID) — cumulative ionizing-radiation effects (oxide trapped charge, parametric drift) & \textbf{Co-60 gamma source} at 1.17\,MeV and 1.33\,MeV. The classical reference test; Co-60 sources are widely available at universities and national labs (24/7 operation is the norm).\\
\bottomrule
\end{tabularx}

\paragraph{Combining 1B + 1C via a single proton test.} Sub-tests 1B and 1C can be merged into a single \textbf{proton irradiation} that simultaneously delivers a NIEL-scaled displacement damage dose and an ionizing dose. The two dose contributions are tracked separately during the test (NIEL\,$\times$\,fluence for DDD; LET\,$\times$\,fluence for TID). Because galactic cosmic rays (GCR) are \textbf{$\sim$90\,\% protons} and solar energetic particles (SEP) are \textbf{$\sim$99\,\% protons}, a proton beam is in fact a closer match to the actual orbital particle environment than separate neutron + gamma irradiations. Many JPL/NASA projects therefore favour the combined-proton path.

\paragraph{Setting the radiation test levels.} Fluence and dose levels are derived from the specified Earth orbit (or planetary destination) and the mission duration; these have to be calculated up-front from the project Radiation Environment Specification (typically using SPENVIS for ESA or OMERE / CREME96 for NASA), with the appropriate Radiation Design Margin (RDM\,$\ge$\,2 in most ESA / NASA practice).

\paragraph{Passive astrophotonic PICs: scaling caveat.} For passive PIC devices — silica AWGs, Si\textsubscript{3}N\textsubscript{4} AWG / MZI / micro-ring structures, ULI-written 3D waveguide chips — the picture is markedly different. These devices have demonstrated excellent tolerance to all three sub-tests at doses far exceeding LEO requirements (see \S\ref{sec:tailoring} rows 13--15, and the Vogl \& Corrielli 2020 ULI data). The radiation campaign therefore reduces to a single confirmation test rather than a full SEE cross-section map, since there are no electronically active junctions to upset. Heavy-ion campaigns retain residual value only to bound any optical-induced effect (transient absorption, colour-centre formation) in the waveguide cladding under high local LET — but the relevance is far lower than for an active III-V component.

\subsection{Mechanical shock and vibration}

Standard GSFC-STD-7000B (GEVS) and ECSS-E-ST-10-03C sinusoidal + random profiles apply, with detailed limits in \S\ref{sec:tailoring} rows 5--6. For an astrophotonic PIC the critical pass/fail criterion is \textbf{no shift in fibre-to-chip alignment} (typical budget: $\Delta$IL\,$\le$\,0.2\,dB), together with integrity of the chip facets and the fibre boot.

\subsection{Thermal cycling}

Standard MIL-STD-883 Method 1010 / ECSS-Q-ST-70-04C cycling (\S\ref{sec:tailoring} row 7). For PICs the test is normally tailored to $-40$\,\textcelsius\ / $+85$\,\textcelsius\ (Cond.\,B) for SiO\textsubscript{2}, Si\textsubscript{3}N\textsubscript{4} and ULI-written devices, with the full $-55$\,\textcelsius\ / $+125$\,\textcelsius\ (Cond.\,C) reserved for components that need to demonstrate compatibility with non-temperature-controlled bays.

\subsection{Working facility set used in practice}

A typical European JPL/NASA-compatible test campaign for an active astrophotonic component can be run on the following facilities, in addition to the formal ESTEC and JPL/GSFC laboratories listed in \S\ref{sec:facilities}:

\begin{itemize}[leftmargin=1.4em]
  \item \textbf{Protons and TID (Co-60) — Helmholtz-Zentrum Berlin, Wannsee site (HZB).} 
  \item \textbf{Heavy ions — RADEF (RADiation Effects Facility), JYFL, Jyväskylä, Finland.} 
  Reference ESA heavy-ion facility for SEE testing.
  \item \textbf{Shock and vibration — Astro- und Feinwerktechnik Adlershof GmbH (Berlin).} See \url{https://www.astrofein.com/umweltsimulation-raumfahrt/}.
  \item \textbf{Thermal cycling — DLR (Deutsches Zentrum für Luft- und Raumfahrt).} Multiple DLR sites support thermal cycling and TVAC; Cologne (DLR Materials and Components) and Bremen are the most common.
\end{itemize}


\section{Optical fibres — companion qualification framework and mission heritage}\label{sec:fibers}
The PIC-oriented flow of \S\ref{sec:tailoring} always rests on an optical fibre interface — chip pigtails, fibre array units (FAUs), spaceflight connectors, and full instrument harnesses. Optical fibres are arguably the most spaceflight-mature photonic component, with continuous mission heritage going back to the 1978 Long Duration Exposure Facility coupons. Their qualification framework — agency standards, defence specifications, telecom reliability documents, and component-level test methods — therefore forms a natural reference set for any PIC packaging or harness design.

\subsection{Background and key documents}

The dominant references in the NASA / ESA fibre-qualification literature are:

\begin{itemize}[leftmargin=1.4em]
  \item \textbf{NASA GSFC Photonics Group.} Melanie Ott and colleagues have authored the principal heritage and qualification review papers, in particular \emph{Space flight applications of optical fiber: 30 years of space flight success} (IEEE AVFOP, 2010) and \emph{Qualification of Commercial Fiber Optic Components for Space Environments} (ESA/NASA Optoelectronics Workshop, 2005). Mission-level qualification reports cover the LRO/LOLA fibre-array bundle assemblies (NTRS 20090006720) and the Mars-environment fibre cable development (ICSO 2010).
  \item \textbf{NASA NEPP.} The Radiation Effects Summary Database on commercially available optical fibre (GSFC Code 562, 2009 and updates), plus LaBel et al.\ on the suitability of fibre-optic links in the space radiation environment (covering MIL-STD-1773 heritage on SAMPEX, XTE, the HST Solid State Recorder, and the AS1773 Fiber-Optic Data Bus).
  \item \textbf{ESA / ESCC photonics activities.} Taugwalder \emph{ESCC Standards, evaluation and qualification of optical fiber connectors for space application} (ICSO 2012/2014) describes the three-level ESCC specification structure (Basic, Generic, Detail) and the qualification of the AVIM / Mini-AVIM connector family — the first ESCC-qualified spaceflight optical connectors. Bringer \emph{ESA Photonic Components Qualifications activities} (ESTEC 2015) covers rad-hard EDFA validation at 1.55\,\textmu m (0.5\,W and 10\,W), TID of doped fibres, thermal vacuum and the ESCC Evaluation Test Plan (ETP) framework. The ESCIES portal hosts the corresponding qualified parts list (QPL / EPPL) and final evaluation reports.
  \item \textbf{Reviews and survey papers.} Girard et al.\ \emph{Recent advances in radiation-hardened fiber-based technologies for space applications} (J.\ Opt., IOP, 2018) gives a comprehensive overview of rad-hard silica and rare-earth-doped fibres, the MIRAS / SMOS optical harness, and the EDRS / Sentinel programs. Berghmans et al., Friebele (NRL) on radiation-induced attenuation in silica fibres, and Ferraro et al.\ (Sensors, 2023) on fibre sensors in harsh and high-radiation aerospace environments complete the picture.
\end{itemize}

\subsection{Qualification standards for spaceflight optical fibres}

The table below compiles the principal standards used to qualify or procure optical fibres, cables, connectors and active fibre assemblies for spaceflight. Coverage spans the fibre itself, the cabled product, terminations and connectors, environmental test methods, workmanship, and reliability. ESCC specifications follow a three-level structure (Basic / Generic / Detail) that differs from IEC; qualified parts appear in the ESCC Qualified Parts List (QPL) or the European Preferred Parts List (EPPL), accessible via the ESCIES portal.

\begin{landscape}
\centering\scriptsize
\setlength{\LTpre}{0pt}\setlength{\LTpost}{0pt}
\begin{longtable}{@{}p{5.5cm} p{3.0cm} p{14.0cm}@{}}
\rowcolor{esablue}\textcolor{white}{\textbf{Document / standard}} & \textcolor{white}{\textbf{Issuing body}} & \textcolor{white}{\textbf{Scope and application}}\\
\endfirsthead
\rowcolor{esablue}\textcolor{white}{\textbf{Document / standard}} & \textcolor{white}{\textbf{Issuing body}} & \textcolor{white}{\textbf{Scope and application}}\\
\endhead

\multicolumn{3}{@{}l}{\cellcolor{section}\textbf{\textcolor{esablue}{A.\ NASA workmanship, parts assurance and design standards}}}\\

NASA-STD-8739.5 (Rev.\,A, Chg.\,2) & NASA & Workmanship Standard for Fiber Optic Terminations, Cable Assemblies, and Installation. Mandatory NASA standard for critical-work fibre-optic hardware: termination, cleaning, inspection, splicing, routing, assembly testing.\\
\rowcolor{rowalt} NASA-STD-8739.6 & NASA & Implementation Requirements for NASA Workmanship Standards: ESD control, environmental control, operator/inspector certification, vision screening.\\
NASA-HDBK-8739.19 & NASA & Companion handbook to the 8739.x workmanship series.\\
\rowcolor{rowalt} NASA EEE-INST-002 & NASA GSFC & Instructions for EEE Parts Selection, Screening, Qualification and Derating — includes guidance for photonic and optoelectronic parts on Class 1/2/3 missions.\\
NASA-STD-8070.1 & NASA & Space Flight System Design and Environmental Test — invokes the workmanship and fibre-optic standards within the broader environmental test envelope.\\
\rowcolor{rowalt} NASA GSFC PPL-21 & NASA GSFC & Preferred Parts List — controlling document for parts selection including fibres and connectors.\\

\multicolumn{3}{@{}l}{\cellcolor{section}\textbf{\textcolor{esablue}{B.\ ESA / ECSS / ESCC standards}}}\\

ESCC 3901 series & ESA / ESCC & Generic Specification for Wires and Cables. ESCC 3901/001 and /002 address detail requirements for space-grade cables, including optical-fibre cables in the high-temperature range.\\
\rowcolor{rowalt} ESCC 3406 series & ESA / ESCC & Detail Specifications for Optical Fibre Connector Sets — AVIM and Mini-AVIM family; first ESCC-qualified connectors for space applications.\\
ESCC-Q-ST-60-05C & ESA / ESCC & Generic Procurement Requirements for Hybrids — applied during EDFA and laser-module qualification.\\
\rowcolor{rowalt} ESCC 23201 & ESA / ESCC & Evaluation Test Programme guidelines for laser diode modules — framework for fibre-coupled laser source qualification.\\
ECSS-Q-ST-70 & ECSS / ESA & Space Product Assurance — Materials, mechanical parts and processes; umbrella document for fibre-cable material selection.\\
\rowcolor{rowalt} ECSS-Q-ST-70-04 & ECSS / ESA & Thermal testing for the evaluation of space materials, processes, mechanical parts and assemblies.\\
ECSS-Q-ST-70-21 & ECSS / ESA & Flammability testing for the screening of space materials.\\
\rowcolor{rowalt} ECSS-Q-ST-70-29 & ECSS / ESA & Determination of offgassing products from materials and assembled articles (toxicity / odour).\\
ECSS-Q-ST-70-02 & ECSS / ESA & Thermal vacuum outgassing test for screening of space materials.\\
\rowcolor{rowalt} ECSS-Q-ST-60 & ECSS / ESA & EEE components general requirements — basis for component qualification flow.\\
ESA / ESCC No.\ 22900 & ESA / ESCC & Total Dose Steady-State Irradiation Test Method — applied to fibres, EDFAs and optoelectronics.\\
\rowcolor{rowalt} ESA / ESCC No.\ 25100 & ESA / ESCC & Single Event Effects Test Method and Guidelines — used for fibre-coupled optoelectronics on SEE-sensitive missions.\\

\multicolumn{3}{@{}l}{\cellcolor{section}\textbf{\textcolor{esablue}{C.\ MIL-PRF / MIL-STD / SAE specifications}}}\\

MIL-PRF-49291 (Rev.\,D, 2024) & U.S.\ DoD (DLA) & Performance Specification: Fiber, Optical (Metric). General specification with slash sheets for single-mode, multimode and radiation-hardened / radiation-resistant fibres (50/125, 62.5/125, 100/140, 200/230, 400/430\,\textmu m, etc.).\\
\rowcolor{rowalt} MIL-PRF-49291/1B, /6, /7, /12 & U.S.\ DoD & Slash sheets for radiation-resistant fibre, including enhanced-performance / aircraft-applications rad-resistant single-mode and multimode fibres.\\
MIL-PRF-85045 & U.S.\ DoD & General Specification for Fiber Optic Cables — environmental, mechanical and optical performance for aerospace cable products.\\
\rowcolor{rowalt} MIL-STD-1773 / SAE AS1773 & U.S.\ DoD / SAE & Fiber Optic Mechanization of a Digital Time-Division Command/Response Multiplex Data Bus. 1\,Mbps and 1/20\,Mbps fibre-optic versions of MIL-STD-1553B; spaceflight heritage on SAMPEX, HST SSR, TRMM, XTE.\\
MIL-STD-38999 (Series III) & U.S.\ DoD & Connectors, electrical, circular — used with MIL-T-29504 fibre-optic termini for spaceflight fibre-optic harnesses (e.g.\ ISS HRDL).\\
\rowcolor{rowalt} MIL-T-29504 & U.S.\ DoD & Termini, Fiber Optic Connector, removable, used inside 38999-style shells.\\
MIL-STD-810 & U.S.\ DoD & Environmental Engineering Considerations and Laboratory Tests — vibration, shock, thermal, humidity, altitude, fluids.\\
\rowcolor{rowalt} MIL-STD-461 & U.S.\ DoD & Requirements for the control of EMI characteristics of subsystems and equipment.\\
MIL-STD-883 (M.\ 1019) & U.S.\ DoD & Test Methods for Microelectronics — Method 1019 (TID) invoked for optoelectronic devices terminating the fibre link.\\
\rowcolor{rowalt} SAE AS5603 & SAE & Aerospace Digital Time-Division Command/Response Multiplex Data Bus, Fiber Optics Mechanization.\\

\multicolumn{3}{@{}l}{\cellcolor{section}\textbf{\textcolor{esablue}{D.\ Industry workmanship and reliability documents}}}\\

IPC / EIA J-STD-001 Space Addendum & IPC / EIA & Space Applications Electronic Hardware Addendum — soldering and electrical assembly companion applied to fibre-coupled electronics.\\
\rowcolor{rowalt} IPC A-640 / IPC D-640 & IPC & Acceptance and Design / Critical-Process Requirements for Optical Fiber, Optical Cable, and Hybrid Wiring Harness Assemblies.\\
Telcordia GR-1221-CORE & Telcordia / Ericsson & Generic Reliability Assurance Requirements for Passive Optical Components — widely used as the commercial baseline before adding space-specific (TID, vacuum, thermal-vacuum) overlays.\\
\rowcolor{rowalt} Telcordia GR-468-CORE & Telcordia & Generic Reliability Assurance Requirements for Optoelectronic Devices (laser diodes, photodiodes, optical receivers, transceivers).\\
Telcordia GR-326-CORE & Telcordia & Generic Requirements for Single-mode Optical Connectors and Jumper Assemblies.\\
\rowcolor{rowalt} Telcordia GR-1209-CORE & Telcordia & Generic Requirements for Passive Optical Components.\\
Telcordia GR-1312-CORE & Telcordia & Generic Requirements for Optically Amplified WDM Networks and Optical Fiber Amplifiers — used for EDFA / booster qualification baseline.\\

\multicolumn{3}{@{}l}{\cellcolor{section}\textbf{\textcolor{esablue}{E.\ Component-level test methods (TIA / IEC / ITU / ASTM / ANSI)}}}\\

TIA / EIA-455 (FOTP) series & TIA & Fiber Optic Test Procedures — foundational test methods cited by NASA-STD-8739.5, MIL-PRF-49291 and ESCC specs. Notable members: FOTP-26 (crush), FOTP-37 (bend), FOTP-62 (macrobend), FOTP-64 (nuclear radiation), FOTP-176 (dimensions).\\
\rowcolor{rowalt} TIA / EIA-455-64 & TIA & Procedure for Measuring Radiation-Induced Attenuation in Optical Fibers and Optical Cables (gamma TID test).\\
TIA / EIA-492 series & TIA & Detail specifications for single-mode (492C) and multimode (492A) fibres (OM1--OM5, OS1/OS2).\\
\rowcolor{rowalt} EIA / TIA-440 & TIA & Fiber Optic Terminology — invoked by NASA-STD-8739.5.\\
IEC 60793-1 / -2 & IEC & Optical Fibres — measurement methods (-1) and product specifications (-2): single-mode (60793-2-50), multimode (60793-2-10).\\
\rowcolor{rowalt} IEC 60794-1 / -2 / -3 & IEC & Optical Fibre Cables — generic specification and detail specs for indoor, outdoor and harsh-environment cabled products.\\
IEC 61300 series & IEC & Fibre optic interconnecting devices and passive components — basic test and measurement procedures (mechanical, environmental, optical).\\
\rowcolor{rowalt} IEC 61753 & IEC & Fibre optic interconnecting devices and passive components performance standard.\\
ITU-T G.652 / G.653 / G.655 / G.657 & ITU-T & Single-mode fibre characteristics (NDSF, DSF, NZ-DSF, bend-insensitive); referenced for telecom-grade fibres used in space links (e.g.\ Corning SMF-28 for SMOS / MIRAS).\\
\rowcolor{rowalt} ITU-T G.651.1 & ITU-T & Characteristics of 50/125\,\textmu m graded-index multimode fibre.\\
ASTM E595 & ASTM & Standard Test Method for Total Mass Loss (TML < 1.0\%) and Collected Volatile Condensable Materials (CVCM < 0.1\%) from Outgassing in a Vacuum Environment — mandatory for all spaceflight fibre jackets and cable materials.\\
\rowcolor{rowalt} ASTM E1559 & ASTM & Standard Test Method for Contamination Outgassing Characteristics of Spacecraft Materials.\\
ANSI Z136.1 / Z136.2 & ANSI & Safe Use of Lasers / Safe Use of Optical Fiber Communication Systems Utilizing Laser Diode and LED Sources — invoked by NASA-STD-8739.5.\\

\end{longtable}
\end{landscape}

\subsection{Space mission heritage with optical fibre}

Optical fibres have flown on space missions for more than four decades. The table below collects representative missions — either as transmission media for data buses and command / telemetry links, or as elements of scientific instruments (laser altimeters, lidars, spectrometers, optical communications terminals). Dates indicate the launch (or first flight) of the relevant element. This list is not exhaustive: many additional missions carry fibre-optic gyroscopes, fibre Bragg-grating strain/temperature sensors on launch vehicles and pressure vessels, and fibre-coupled laser sources within instrument optical trains.

\begin{landscape}
\centering\scriptsize
\setlength{\LTpre}{0pt}\setlength{\LTpost}{0pt}
\begin{longtable}{@{}p{5.0cm} p{2.3cm} p{2.8cm} p{12.5cm}@{}}
\rowcolor{esablue}\textcolor{white}{\textbf{Mission / platform}} & \textcolor{white}{\textbf{Year}} & \textcolor{white}{\textbf{Agency / lead}} & \textcolor{white}{\textbf{Optical-fibre role and notes}}\\
\endfirsthead
\rowcolor{esablue}\textcolor{white}{\textbf{Mission / platform}} & \textcolor{white}{\textbf{Year}} & \textcolor{white}{\textbf{Agency / lead}} & \textcolor{white}{\textbf{Optical-fibre role and notes}}\\
\endhead

LDEF (Long Duration Exposure Facility) & 1984 (rec.\ 1990) & NASA & One of the earliest space environment exposure tests including optical-fibre coupons; baseline for radiation and atomic-oxygen response of early fibres.\\
\rowcolor{rowalt} SAMPEX & 1992 & NASA / SMEX & First operational MIL-STD-1773 fibre-optic data bus on a NASA SMEX spacecraft (1\,Mbps, 850\,nm).\\
X-ray Timing Explorer (XTE / RXTE) & 1995 & NASA & MIL-STD-1773 fibre-optic data link heritage; demonstration of fibre bus reliability over mission life.\\
\rowcolor{rowalt} Mars Global Surveyor — MOLA & 1996 & NASA & Mars Orbiter Laser Altimeter; fibre-coupled receiver heritage that fed forward to MLA and LOLA.\\
TRMM (Tropical Rainfall Measuring Mission) & 1997 & NASA / JAXA & MIL-STD-1773, 1\,Mbps, 850\,nm fibre-optic command/data bus.\\
\rowcolor{rowalt} Hubble Space Telescope — Servicing Mission upgrades & 1997 onward & NASA & Solid State Recorder upgrade used MIL-STD-1773 fibre link (850\,nm, 1\,Mbps); fibre introduced via servicing missions.\\
Photonics Space Experiment (PSE) & 1998 & Boeing / USAF & Boeing-led demonstration of fibre-optic data link components in LEO.\\
\rowcolor{rowalt} International Space Station — HRDL & 2001 onward & NASA / Boeing & High Rate Data Link: 1300\,nm graded-index multimode fibre-optic bus on the ISS using MIL-STD-38999 shells with MIL-T-29504 fibre termini.\\
ICESat — GLAS & 2003 & NASA / GSFC & Geoscience Laser Altimeter System using Diamond AVIM optical-fibre connectors and fibre-coupled detector chains.\\
\rowcolor{rowalt} MESSENGER — Mercury Laser Altimeter (MLA) & 2004 & NASA / APL / GSFC & Fibre-coupled receiver / transmitter pathways; AVIM connectors. Demonstrated two-way laser ranging across 24 million km between MLA and GGAO.\\
Shuttle Return-to-Flight (STS-114, 121) & 2005--2006 & NASA & High-definition heat-tile inspection camera with optical fibre array and AVIM connectors.\\
\rowcolor{rowalt} SMOS — MIRAS instrument & 2009 & ESA & Soil Moisture and Ocean Salinity: MIRAS optical harness distributes a 56\,MHz clock to 69 L-band receivers over Corning SMF-28 single-mode fibre at 1300\,nm.\\
Lunar Reconnaissance Orbiter — LOLA \& LR & 2009 & NASA / GSFC & Lunar Orbiter Laser Altimeter and Laser Ranging application; custom optical-fibre array bundle assemblies built and qualified at GSFC using 200/220\,\textmu m step-index FI-series fibre and modified AVIM connectors.\\
\rowcolor{rowalt} Express Logistics Carrier (ELC) — ISS & 2009--2011 & NASA / GSFC & ISS optical fibre subsystem components carried on the ELC platform; integration with HRDL.\\
Mars Science Laboratory — ChemCam & 2011 & NASA / CNES & Curiosity rover laser-induced breakdown spectroscopy: optical fibre routes plasma emission from the mast unit to the body-mounted spectrometers.\\
\rowcolor{rowalt} LADEE — LLCD (Lunar Laser Communications Demonstration) & 2013 & NASA / MIT-LL & Free-space laser communications demonstration to and from the Moon at up to 622\,Mbps; fibre-coupled space-terminal optics.\\
Sentinel-1/2/3 (Copernicus) & 2014 onward & ESA / EU & Optical inter-satellite links via EDRS; on-board fibre harness for sensor data routing.\\
\rowcolor{rowalt} EDRS — European Data Relay System & 2016 onward & ESA / Airbus & Inter-satellite optical communications laser terminals using EDFA-amplified 1064 / 1550\,nm links; relays Sentinel data to ground.\\
NASA ISS — Fiber Optic Production / Space Fibers & 2017, 2019, ongoing & NASA / FOMS / Made In Space & Microgravity manufacturing of ZBLAN fluoride fibres on the ISS — fibre as the \emph{product} rather than the link.\\
\rowcolor{rowalt} ICESat-2 — ATLAS & 2018 & NASA / GSFC & Advanced Topographic Laser Altimeter System: six-beam micropulse 532\,nm photon-counting lidar; fibre-coupled optical filter assemblies and laser transmitter chain qualified at GSFC.\\
BepiColombo — MERTIS / ISA harness elements & 2018 & ESA / JAXA & Fibre-optic elements within instrument calibration and signal routing in the Mercury Planetary Orbiter; radiation-hardened components selected for high-temperature environment.\\
\rowcolor{rowalt} Solar Orbiter — METIS / SPICE harness & 2020 & ESA / NASA & Fibre-optic links inside coronagraph and EUV spectral imager instruments.\\
Mars 2020 / Perseverance — SuperCam & 2020 & NASA / CNES / IRAP & 5.8\,m optical-fibre bundle routes light from the rover-mast telescope to body-mounted spectrometers (UV/VIO/VIS/NIR + Raman); flight qualification including Mars thermal cycling.\\
\rowcolor{rowalt} JWST & 2021 & NASA / ESA / CSA & Used copper twisted-pair (SpaceWire) rather than optical fibre for the main interconnect — listed here as a counter-example illustrating that fibre is not yet ubiquitous on every flagship.\\
LCRD — Laser Communications Relay Demonstration & 2021 & NASA & GEO laser-comm relay; fibre-amplified 1550\,nm transmitters and fibre-coupled receivers.\\
\rowcolor{rowalt} ARTEMIS / Orion fibre-optic instrumentation & 2022 onward & NASA & Fibre-optic strain and temperature sensors and data links on Orion test articles.\\
TBIRD — TeraByte InfraRed Delivery & 2022 & NASA / MIT-LL & 200\,Gbps optical downlink CubeSat demonstration using fibre-coupled transceiver technology.\\
\rowcolor{rowalt} Psyche — DSOC payload & 2023 & NASA / JPL & Deep Space Optical Communications technology demonstration; fibre-coupled photon-counting detector and laser transmitter components.\\

\end{longtable}
\end{landscape}

\subsection{Additional heritage: FOG-based IRUs, FBG sensor instrumentation and fibre-coupled instrument lasers}

The chronological table above lists missions where the optical fibre is the \emph{transmission medium} for data or instrument signals. Three further mission categories carry optical fibre in other functional roles and complete the heritage picture:

\begin{enumerate}[leftmargin=1.6em,nosep]
  \item Inertial Reference Units (IRUs) built around \textbf{fibre-optic gyroscopes (FOGs)} or, in early deep-space heritage, \textbf{hybrid / hemispherical resonator gyroscopes (HRGs)} that paved the way for all-photonic rotation sensing.
  \item \textbf{Fibre Bragg-grating (FBG) strain and temperature sensors} on launch vehicles, pressure vessels and spacecraft structures.
  \item \textbf{Fibre-coupled laser sources} forming part of an instrument's optical train (typically fibre-pigtailed pump diodes or fibre-delivered seed lasers feeding a free-space gain stage).
\end{enumerate}

\begin{landscape}
\centering\scriptsize
\setlength{\LTpre}{0pt}\setlength{\LTpost}{0pt}
\begin{longtable}{@{}p{4.0cm} p{5.0cm} p{1.8cm} p{2.5cm} p{9.0cm}@{}}
\rowcolor{esablue}\textcolor{white}{\textbf{Category}} & \textcolor{white}{\textbf{Mission / platform}} & \textcolor{white}{\textbf{Year}} & \textcolor{white}{\textbf{Agency / lead}} & \textcolor{white}{\textbf{Role and notes}}\\
\endfirsthead
\rowcolor{esablue}\textcolor{white}{\textbf{Category}} & \textcolor{white}{\textbf{Mission / platform}} & \textcolor{white}{\textbf{Year}} & \textcolor{white}{\textbf{Agency / lead}} & \textcolor{white}{\textbf{Role and notes}}\\
\endhead

\multicolumn{5}{@{}l}{\cellcolor{section}\textbf{\textcolor{esablue}{F.\ Fibre-optic / photonic gyroscope IRU heritage}}}\\

Photonic IRU — deep space, early heritage & Galileo (Jupiter orbiter) & 1989 & NASA / JPL & Early deep-space IRU heritage with Dry Tuned Gyro (DTG) sensors; predecessor configuration to the photonic IRUs that followed on Cassini and later missions.\\
\rowcolor{rowalt} Photonic IRU — deep space (HRG) & Cassini-Huygens & 1997 & NASA / JPL / ESA & JPL Inertial Reference Unit built around Litton (Delco) Hemispherical Resonator Gyroscopes — the first deep-space mission to use an all-photonic, non-spinning-mass attitude-rate sensor; foundational heritage for later FOG-based IRUs.\\
FOG IRU — commercial GEO & Honeywell ``Spirit'' IFOG IRU & 2011+ & Honeywell SSD (USA) & Interferometric Fiber-Optic Gyroscope IRU developed for pointing commercial communications satellites; complements the earlier MIMU mechanical heritage.\\
\rowcolor{rowalt} FOG IRU — Earth observation & Sentinel-1 / Sentinel-2 / Sentinel-3 (Copernicus) & 2014+ & ESA / Airbus / Exail (ex-iXBlue) & Airbus / Exail Astrix-series FOG IRU on the operational Copernicus Sentinels; representative European space-grade FOG flight heritage.\\
FOG IRU — interplanetary & ExoMars TGO; BepiColombo MPO; JUICE & 2016 / 2018 / 2023 & ESA / Airbus / NG / Exail & Northrop Grumman SIRU (FOG) and Astrix-200 / NS FOG IRUs flown on flagship ESA missions to Mars, Mercury and the Jovian system; demonstrate radiation-tolerant FOG technology beyond Earth orbit.\\
\rowcolor{rowalt} FOG IMU — smallsats / CubeSats & Numerous commercial Earth-observation constellations (Planet, BlackSky, etc.) and university CubeSats & 2014+ & Commercial & Adoption of COTS FOG-based IMUs (typically miniature LN-200-class) for attitude determination on smallsat platforms.\\

\multicolumn{5}{@{}l}{\cellcolor{section}\textbf{\textcolor{esablue}{G.\ Fibre Bragg-grating (FBG) sensor instrumentation}}}\\

FBG strain network on reusable spacecraft & X-38 Crew Return Vehicle (composite skin instrumentation) & ca.\ 2000 (test article; mission cancelled) & NASA / DLR / industry & Early flight-representative demonstration of an FBG strain-sensor network for thermo-mechanical monitoring of composite re-entry structures (Frövel et al., SPIE 2000).\\
\rowcolor{rowalt} FBG strain / temperature instrumentation on cryogenic stages and COPVs & Ariane 5 / Vega-C upper stages and Composite Overwrapped Pressure Vessels & various & ESA / ArianeGroup / Avio / industry & Surface-mounted FBG sensor arrays for structural health monitoring (SHM) of cryogenic tank skins, COPVs, and interstage composite assemblies; large body of dedicated qualification work since the 2000s.\\
FBG in-orbit demo & INTA Fiber-Optic Sensor Demonstrator (FSD) & 2018 & INTA (ES) / ESA & In-orbit demonstration of FBG temperature and pressure sensor heads on a small satellite payload; documented in López-Heredero et al., ICSO 2018 (Proc.\ SPIE 11180).\\
\rowcolor{rowalt} FBG / fibre sensors on ISS external payloads & ISS — various external instrumentation packages & 2010s+ & NASA / ESA / JAXA & FBG-based temperature and strain monitoring on external pallets, antenna structures and thermal-protection samples; complements the HRDL fibre-data heritage on the same platform.\\
FBG strain sensors on launch-vehicle test articles & Florida Tech / NASA Kennedy SHM research articles; reusable-launch-vehicle SHM activities & 2015+ & NASA / industry / academia & Ground- and flight-representative campaigns demonstrating FBG arrays for dynamic-strain measurement during separation events, re-flight certification of reusable composite stages, and vibration control (van der Veek et al., \emph{Sensors} 25, 2025).\\

\multicolumn{5}{@{}l}{\cellcolor{section}\textbf{\textcolor{esablue}{H.\ Fibre-coupled laser sources within instrument optical trains}}}\\

Fibre-coupled pump diodes — Earth-observation lidar & CALIPSO / CALIOP & 2006 & NASA / CNES & Polarisation-sensitive 532 / 1064\,nm cloud-aerosol lidar based on diode-pumped Nd:YAG laser; pump-diode coupling and seed delivery use fibre interfaces.\\
\rowcolor{rowalt} Fibre-coupled diode pumps — wind lidar & Aeolus / ALADIN & 2018 & ESA / Airbus DS & First spaceborne wind lidar; 355\,nm transmitter based on a frequency-tripled diode-pumped Nd:YAG laser, the high-power pump diode arrays being qualified for fibre-coupled delivery to the Nd:YAG gain medium.\\
Fibre-coupled diode pumps — atmospheric lidar & EarthCARE / ATLID & 2024 & ESA / JAXA / Airbus DS & 355\,nm backscatter atmospheric lidar; same diode-pumped Nd:YAG architecture as ALADIN, leveraging the qualified fibre-coupled high-power laser-diode array heritage.\\
\rowcolor{rowalt} Fibre-coupled calibration / metrology laser & BepiColombo MERTIS, Solar Orbiter SPICE, JUICE GALA, ICESat-2 ATLAS receivers & 2018+ & ESA / NASA / DLR / GSFC & Internal calibration, metrology and reference-laser delivery via short fibre links within the instrument optical train (in addition to the science-photon paths already listed in the chronological heritage table).\\

\end{longtable}
\end{landscape}

\subsection{Typical qualification test categories for spaceflight optical fibres}

A spaceflight optical fibre, cable, or assembly qualification campaign typically combines test blocks from several of the standards listed above. The principal categories common to NASA, ESCC and DoD flows are:

\begin{itemize}[leftmargin=1.4em]
  \item \textbf{Construction analysis and visual inspection.} Destructive Physical Analysis (DPA); end-face geometry (interferometric) per NASA-STD-8739.5; ferrule and connector dimensions per IEC 61300 / TIA-455-176.
  \item \textbf{Optical performance.} Insertion loss, return loss, attenuation vs.\ wavelength (TIA-455-78), modal bandwidth (multimode), polarization-mode dispersion (PMD).
  \item \textbf{Mechanical — vibration and shock.} Random / sine vibration (qualification levels typically 15--40\,g rms), pyro shock, mechanical shock (TIA-455-14, MIL-STD-810, ECSS-E-ST-10-03).
  \item \textbf{Mechanical — cable performance.} Crush (FOTP-26), impact (FOTP-25), tensile load (FOTP-33), twist (FOTP-85), bend (FOTP-37), macrobend (FOTP-62).
  \item \textbf{Thermal cycling and thermal vacuum.} Vacuum bake-out per ASTM E595 / E1559; thermal cycling typically $-55$\,\textcelsius\ to $+85$\,\textcelsius\ (commercial) or $-65$\,\textcelsius\ to $+125$\,\textcelsius\ (high-temperature ESCC 3901/001); thermal vacuum cycling per ECSS-Q-ST-70-04.
  \item \textbf{Radiation — total ionising dose (TID).} Gamma \textsuperscript{60}Co irradiation per TIA/EIA-455-64, ESA/ ESCC 22900, MIL-STD-883 Method 1019; dose rates tailored to mission orbit (LEO, MEO, GEO, interplanetary).
  \item \textbf{Radiation — proton and heavy-ion (SEE).} Single Event Effects on the optoelectronics terminating the fibre link (ESA/ESCC 25100).
  \item \textbf{Outgassing and materials.} TML < 1.0\%, CVCM < 0.1\% per ASTM E595; flammability per ECSS-Q-ST-70-21; offgassing per ECSS-Q-ST-70-29.
  \item \textbf{Endurance / life.} Damp heat (85\,\textcelsius\ / 85\,\%RH), high-temperature storage, low-temperature storage, mating cycle endurance for connectors (typically 500--2000 cycles).
  \item \textbf{Workmanship and process control.} NASA-STD-8739.5 / 8739.6, IPC J-STD-001 Space Addendum, operator certification.
\end{itemize}

\subsection{Document links (2026)}

\begin{itemize}[leftmargin=1.4em,nosep]
  \item NASA standards portal — \url{https://standards.nasa.gov}
  \item NASA NEPP (Electronic Parts and Packaging) — \url{https://nepp.nasa.gov}
  \item NASA GSFC Photonics Group documents — \url{https://photonics.gsfc.nasa.gov/photonics/}
  \item ESA / ESCC component portal (ESCIES) — \url{https://escies.org}
  \item ESA Space Components information and ESCC specifications — \url{https://spacecomponents.org}
  \item European Cooperation for Space Standardization (ECSS) — \url{https://ecss.nl}
  \item U.S.\ DoD military spec repository (DLA Land \& Maritime) — \url{https://landandmaritimeapps.dla.mil}
  \item Telcordia GR documents — \url{https://telecom-info.njdepot.ericsson.net}
  \item TIA standards — \url{https://tiaonline.org}
  \item IEC technical committee TC86 (Fibre optics) — \url{https://www.iec.ch/tc86}
  \item ITU-T G-series recommendations — \url{https://www.itu.int/rec/T-REC-G}
\end{itemize}

\section{Key test facilities}\label{sec:facilities}
\subsection{European facilities}
\begin{itemize}[leftmargin=1.4em]
  \item \textbf{ESTEC Test Centre, Noordwijk (NL)} — Large Space Simulator (LSS), HYDRA multi-axis shaker, acoustic chamber, EMC Maxwell chamber.
  \item \textbf{ESA-ESTEC Optics \& Opto-Electronics Lab (OOEL) and Materials \& Electrical Components Lab (M\&ECL)} — Co-60 facility (ISO/IEC 17025), photonic component characterisation; ESA SCC certification authority.
  \item \textbf{Centre Spatial de Liège, CSL (BE)} — FOCAL cryogenic vacuum chambers used for JWST NIRSpec, MTG, PLATO; specialised in optical instrument TVAC.
  \item \textbf{UCL Cyclotron Resource Centre, Louvain-la-Neuve (BE)} — protons (LIF, 10--75\,MeV), heavy ions (HIF, CYCLONE), neutrons (NIF), Co-60 (GIF). Reference EU radiation facility for ESA.
  \item \textbf{PSI Proton Irradiation Facility, Villigen (CH)} — protons up to 230\,MeV.
  \item \textbf{RADEF, JYFL, Jyväskylä (FI)} — heavy ions and protons.
  \item \textbf{IABG, Ottobrunn (DE)} — vibration, TVAC, MFSA magnetic field simulation.
  \item \textbf{Tyndall National Institute, Cork (IE)} — ESA Microelectronics Technology Support Lab.
  \item \textbf{Helmholtz-Zentrum Berlin (HZB), Wannsee site (DE)} — proton beam and Co-60 gamma source (24/7) for TID and combined TID+DDD proton testing.
  \item \textbf{Astro- und Feinwerktechnik Adlershof GmbH (Berlin, DE)} — shock and vibration testing for spaceflight components, \url{https://www.astrofein.com/umweltsimulation-raumfahrt/}.
  \item \textbf{Fraunhofer IOF (Jena), IZM (Berlin), CSEM (CH), CEA-LETI (Grenoble)} — packaging, photonics characterisation, hermetic packaging.
  \item \textbf{Alter Technology, Airbus DS, ONERA (FR), DLR (DE — thermal cycling and TVAC at Cologne / Bremen), AIT (AT), INTA (ES)} — independent testing under ESA contracts.
\end{itemize}

\subsection{US facilities}
\begin{itemize}[leftmargin=1.4em]
  \item \textbf{NASA Goddard Space Flight Center (GSFC)} — Environmental Test \& Integration Branch, Photonics Group (Code 562), REAG (Co-60), NEPP program lead.
  \item \textbf{NASA Jet Propulsion Laboratory (JPL)} — 25-foot and 10-foot Space Simulators (TVAC), vibration tables, Microdevices Lab, IBA Industrial Dynamitron (3\,MeV electrons), Co-60 irradiator.
  \item \textbf{NASA Johnson Space Center (JSC)} — Thermal Vacuum Chambers A and B (Chamber A used for JWST OTIS).
  \item \textbf{NASA Glenn / Plum Brook Station} — Space Power Facility (world's largest TVAC chamber).
  \item \textbf{NASA Marshall Space Flight Center (MSFC)} — vibration, EMC, thermal vacuum.
  \item \textbf{NASA Space Radiation Laboratory (NSRL) at BNL} — heavy ions and protons up to GeV energies.
  \item \textbf{Lockheed Martin Space Plasma and Radiation Center (SPARC) Lab and Parts, Materials, \& Processes (PMP) Labs} –   DPA, Optical performance (tabletop, in situ and in vacuum, UV-Vis-NIR-IR), Mechanical vibration and shock, Mechanical cable performance, Thermal Cycling and Thermal vacuum, TID (gamma), Outgassing (ASTM E595), Endurance / life (environmental exposures and aging), electron exposures 0.2 to 100 keV, proton exposures 20-200 keV, spectrally matched AM0 Solar simulators.
  \item \textbf{Texas A\&M University Cyclotron Institute} — heavy ions, primary US SEE facility.
  \item \textbf{Lawrence Berkeley National Lab 88-Inch Cyclotron} — heavy ion SEE testing.
  \item \textbf{UC Davis Crocker Nuclear Lab, MGH Burr Center, Indiana Univ.\ Cyclotron} — proton irradiations.
  \item \textbf{Sandia National Lab, NRL, Aerospace Corp.} — RHA, ESD, reliability.
  \item \textbf{NIST Boulder, MIT Lincoln Lab.} — optical metrology and photonic device qualification.
\end{itemize}


\section{Results and discussion}\label{sec:results}
This review has assembled, for the first time in a single place, the elements
needed to qualify an astrophotonic photonic integrated circuit (PIC) for
spaceflight: a 19-step master qualification flow with its applicable
standards and test facilities (\S\ref{sec:tailoring}); a project-specific
tailoring matrix that bounds the flow between a LEO smallsat technology
demonstrator and an HWO-class L2 flagship (\S\ref{sec:tailoring}); a
TRL-versus-test-coverage roadmap that maps each activity onto the NASA/ESA
readiness levels and review gates (\S\ref{sec:trl},
Fig.~\ref{fig:trlcoverage}); a survey of UV, visible and near-infrared
material platforms with their spectral reach, maturity and flight heritage
(\S\ref{sec:platforms}, Figs.~\ref{fig:wavelength},
\ref{fig:sciencecase} and~\ref{fig:landscape}); a consolidated
radiation-effects summary (\S\ref{sec:radsummary},
Table~\ref{tab:radsummary}); a set of JPL/NASA practitioner notes
(\S\ref{sec:practitioner}); and four decades of optical-fibre mission
heritage (\S\ref{sec:fibers}). Read together, these elements support a
small number of conclusions and, more importantly, a reusable
qualification template that ESA and NASA programmes can adopt directly. Tables~\ref{tab:template}, \ref{tab:procedures} and~\ref{tab:gaps} constitute the original contribution of this work --- the proposed PIC-SQT template, its expansion into exact test procedures, and the analysis of qualification processes not yet covered by existing standards; the remaining tables compile existing ESA/NASA standards and flight heritage.

\subsection{Principal findings}
\begin{enumerate}[leftmargin=1.6em]
  \item \textbf{Tailoring is mandatory, not optional.} Astrophotonic PICs do
  not yet belong to a closed family of qualified EEE/optoelectronic parts.
  Both the ESA ECSS-Q-ST-60 system and the NASA EEE-INST-002 /
  NASA-STD-8739.11 framework explicitly permit tailoring of the
  qualification flow for novel devices; the
  19-step flow of \S\ref{sec:tailoring} is therefore best regarded as a
  superset from which each project subtracts and adjusts.

  \item \textbf{The radiation burden is concentrated in the active building
  blocks.} Passive low-index-contrast platforms (silica-on-silicon,
  Si\textsubscript{3}N\textsubscript{4}, ULI-written glass) are essentially
  radiation-immune at LEO/GEO total-ionising-dose and proton
  fluences~\cite{piacentini2020,yin2021}, whereas Ge-on-Si photodiodes,
  p--i--n modulators and III--V active devices dominate the
  displacement-damage and single-event
  budgets~\cite{mao2024,terrasanta2025}. Qualification effort, beam time and
  radiation-design margin should be allocated accordingly --- a single
  confirmation campaign for passive cores, a full SEE/TID/DDD matrix for
  active devices.

  \item \textbf{Two concrete space-qualification campaigns anchor the
  field.} The ULI-glass campaign of Piacentini and
  co-workers~\cite{piacentini2020} and the silicon-photonics campaign of
  Mao and co-workers~\cite{mao2024} are the two most complete open-literature
  demonstrations and serve as worked examples for the template proposed
  below. UV-transparent platforms (ALD alumina, AlF\textsubscript{3},
  AlN-on-sapphire) remain at TRL~3--4 with strong materials demonstrations
  but no published space qualification (Fig.~\ref{fig:landscape}).

  \item \textbf{Mission class drives rigour more than it drives the test
  set.} The LEO Class~D and HWO Class~A profiles of \S\ref{sec:tailoring}
  share essentially the same physical test list; they differ chiefly in
  sample sizes, radiation-design margin, the depth of documentation, and the
  stability (sub-mK, sub-pm) demanded by the $10^{-10}$ contrast budget.
  Programmatic rigour and stability verification, not exotic new tests, are
  the principal cost differentials.

  \item \textbf{The gating gaps are packaging, UV-platform RHA, cryogenics,
  hybrid integration and space-grade PDKs.} These five items, identified
  consistently across the platform survey and the 2023 Astrophotonics
  Roadmap~\cite{jovanovic2023} (\S\ref{sec:platform_outlook}), are what
  currently prevent visible/NIR PICs from crossing the TRL-6 flight gate and
  UV PICs from leaving the laboratory.
\end{enumerate}

\subsection{A proposed space-qualification template for astrophotonic PICs (PIC-SQT)}
The central practical outcome of this review is the qualification
\emph{template} of Table~\ref{tab:template}. It re-organises the 19-step
master flow of \S\ref{sec:tailoring} into seven sequential phases, each with
an explicit entry/exit gate keyed to the TRL ladder of \S\ref{sec:trl}, the
governing ESA and NASA standards, the data products that constitute the exit
evidence, and the tailoring to be applied for the two bounding mission
classes. The template is deliberately device-agnostic at the top level and
becomes specific through two tailoring decisions taken at Phase~0: (i)
whether the device is \emph{passive} or \emph{active}, which sets the depth
of the radiation campaign (Phase~4); and (ii) the \emph{mission class}
(LEO Class~D $\leftrightarrow$ HWO Class~A), which sets sample sizes,
margins and documentation. Used this way, the same template produces a
single-string COTS-with-uprating flow for a CubeSat demonstrator and a
fully traceable Class~A flow for a flagship, without changing its structure.

\begin{landscape}
\centering\scriptsize
\setlength{\LTpre}{0pt}\setlength{\LTpost}{0pt}
\begin{longtable}{@{}p{2.3cm} p{3.5cm} p{1.9cm} p{4.4cm} p{4.0cm} p{4.6cm}@{}}
\caption{Proposed space-qualification template for astrophotonic PICs
(PIC-SQT). The seven phases re-organise the 19-step master flow of
\S\ref{sec:tailoring}; entry/exit gates are keyed to the TRL roadmap of
\S\ref{sec:trl}. Standards are indicative; the project Environmental and
Radiation Environment Specifications remain authoritative.}\label{tab:template}\\
\rowcolor{esablue}\textcolor{white}{\textbf{Phase / TRL gate}} &
\textcolor{white}{\textbf{Activities (master-table steps)}} &
\textcolor{white}{\textbf{Exit TRL}} &
\textcolor{white}{\textbf{Governing standards (ESA / NASA)}} &
\textcolor{white}{\textbf{Exit data products}} &
\textcolor{white}{\textbf{Tailoring: LEO Class~D / HWO Class~A}}\\
\endfirsthead
\rowcolor{esablue}\textcolor{white}{\textbf{Phase / TRL gate}} &
\textcolor{white}{\textbf{Activities (master-table steps)}} &
\textcolor{white}{\textbf{Exit TRL}} &
\textcolor{white}{\textbf{Governing standards (ESA / NASA)}} &
\textcolor{white}{\textbf{Exit data products}} &
\textcolor{white}{\textbf{Tailoring: LEO Class~D / HWO Class~A}}\\
\endhead

\textbf{0. Classification \& requirements} (entry TRL 3) &
Device class (passive vs active); mission \& orbit; Environmental \&
Radiation Environment Specs; tailoring decisions. &
TRL 3$\rightarrow$4 &
ECSS-E-ST-10-04C Rev.1; ECSS-Q-ST-60C Rev.4 / -60-13C; NASA-STD-8719.14C; EEE-INST-002. &
Tailoring matrix; declared device class; requirement \& margin baseline. &
\cellcolor{leo}Single-string COTS-with-uprating; PAL~4 / Class~3 acceptable. \cellcolor{hwo}Class~S/1 screening; full traceability from the outset.\\

\rowcolor{rowalt}\textbf{1. Pre-screening \& baseline} (steps 1--4) &
Visual/dimensional; optical/electrical baseline; hermeticity; PIND. &
TRL 4$\rightarrow$5 &
MIL-STD-883 (2009/2010/1014/2020); ECSS-Q-ST-60-05C Rev.1; Telcordia GR-468-CORE. &
As-built baseline dataset (IL, RL, PDL, spectral map); seal record. &
\cellcolor{leo}Sample-based; hermeticity optional for coated SiPh. \cellcolor{hwo}100\% inspection; mandatory hermeticity + RGA.\\

\textbf{2. Environmental} (steps 5--10) &
Vibration; shock/accel.; thermal cycling; TVAC; damp heat; outgassing. &
TRL 5$\rightarrow$6 &
GSFC-STD-7000B GEVS; ECSS-E-ST-10-03C Rev.1; ECSS-Q-ST-70-02C/-04C; ASTM~E595. &
Protoflight/qual test reports; $\Delta$IL \& $\Delta\lambda$ records; TML/CVCM. &
\cellcolor{leo}Acceptance or protoflight; launcher-tailored levels. \cellcolor{hwo}Full qual + protoflight; cryogenic cycling for IR PICs.\\

\rowcolor{rowalt}\textbf{3. Reliability \& life} (steps 11--12) &
HTOL; burn-in (active devices). &
TRL 6 &
MIL-STD-883 (1005/1015); Telcordia GR-468-CORE \S5--6. &
Arrhenius-extrapolated life; FIT/failure-rate projection. &
\cellcolor{leo}1000\,h, small sample (LTPD~20). \cellcolor{hwo}2000\,h, 22-device (LTPD~10), full-life extrapolation.\\

\textbf{4. Radiation / RHA} (steps 13--16) &
TID (Co-60); DDD (protons); SEE (heavy ions); UV/VUV --- depth set by Phase~0 class. &
TRL 6 &
ECSS-Q-ST-60-15C Rev.1; ESCC 22900/25100; MIL-STD-883 M.1019; ASTM~E722~\cite{piacentini2020,mao2024}. &
TID/DDD/SEE cross-sections; RDM-justified dose; pass/fail vs spec. &
\cellcolor{leo}Single confirmation test for passive cores; LET~37. \cellcolor{hwo}Full matrix; LET~80 SEL-immunity; multi-energy DDD.\\

\rowcolor{rowalt}\textbf{5. ESD, EMC \& final acceptance} (steps 17--19) &
ESD (HBM/CDM); EMC/EMI; end-of-line acceptance. &
TRL 6$\rightarrow$7 &
MIL-STD-883 M.3015; JEDEC JS-001/002; MIL-STD-461G; ECSS-Q-ST-10-09C Rev.1 / -20C Rev.2. &
ESD class; EMC report; Qualification Test Report; DML/DCL. &
\cellcolor{leo}Class~1A; tailored RE/CE; project waivers permitted. \cellcolor{hwo}Class~2 + CDM; full GEVS suite; IRB sign-off.\\

\textbf{6. Flight demonstration \& heritage} (steps 20--22) &
In-flight demonstration; operational use; heritage capture. &
TRL 7$\rightarrow$9 &
NPR 8735.1; mission assurance; ESA in-orbit-demonstration practice. &
On-orbit performance record; lessons-learned; updated heritage. &
\cellcolor{leo}LEO smallsat is itself the TRL-7 vehicle. \cellcolor{hwo}Heritage inherited from LEO demos; closes TRL~8--9 on the flagship.\\

\end{longtable}
\end{landscape}

\subsection{Test procedures of governing standards}
Table~\ref{tab:procedures} expands the ``Governing standards'' column of the
PIC-SQT (Table~\ref{tab:template}, Phases~0--6) into the \emph{exact} test
methods and procedures relevant to astrophotonic PICs, together with the
PIC-specific conditions and acceptance criteria drawn from the master
qualification flow of \S\ref{sec:tailoring}. Where a phase rests on a
framework or environment-definition document rather than a bench procedure
(notably Phases~0 and~6), this is stated explicitly.

\begin{landscape}
\centering\scriptsize
\setlength{\LTpre}{0pt}\setlength{\LTpost}{0pt}
\begin{longtable}{@{}p{3.7cm} p{6.0cm} p{10.6cm}@{}}
\caption{Exact test procedures for astrophotonic PICs extracted from the
governing standards of each PIC-SQT phase (Table~\ref{tab:template}).
Method numbers and conditions are consolidated from the master qualification
flow of \S\ref{sec:tailoring}; project Environmental and Radiation
Environment Specifications remain authoritative for the numerical
levels.}\label{tab:procedures}\\
\rowcolor{esablue}\textcolor{white}{\textbf{Governing standard}} &
\textcolor{white}{\textbf{Exact test method / procedure}} &
\textcolor{white}{\textbf{PIC-relevant condition \& acceptance}}\\
\endfirsthead
\rowcolor{esablue}\textcolor{white}{\textbf{Governing standard}} &
\textcolor{white}{\textbf{Exact test method / procedure}} &
\textcolor{white}{\textbf{PIC-relevant condition \& acceptance}}\\
\endhead

\multicolumn{3}{@{}l}{\cellcolor{section}\textbf{\textcolor{esablue}{Phase 0 --- Classification \& requirements (framework / environment definition; no bench procedure)}}}\\
ECSS-E-ST-10-04C & Space environment specification (radiation, thermal, vacuum, atomic-oxygen models) & Source of the Radiation Environment Spec and thermal/vacuum levels that set all downstream test magnitudes (SPENVIS / OMERE inputs); no coupon test.\\
\rowcolor{rowalt} ECSS-Q-ST-60C Rev.4 (\S6.3) & EEE general requirements; component class; visual-acceptance criteria & Defines Class~1/2/3 and the \S6.3 visual criteria later applied to PIC die, facets and package.\\
ECSS-Q-ST-60-13C & Commercial EEE uprating \& evaluation requirements & Route for COTS PIC uprating (LEO Class-D); defines the Evaluation Test Plan.\\
\rowcolor{rowalt} NASA EEE-INST-002 (+Add.~1) & Parts selection, screening \& derating tables & Screening matrices and derating factors applied to active PIC drivers, lasers and photodiodes.\\
NASA-STD-8719.14C & Orbital-debris / end-of-life assessment & Programmatic; constrains materials and passivation; no PIC bench test.\\

\multicolumn{3}{@{}l}{\cellcolor{section}\textbf{\textcolor{esablue}{Phase 1 --- Pre-screening \& baseline (steps 1--4)}}}\\
MIL-STD-883 Method 2009 & External visual inspection & 100\% inspection of die, waveguide facets, fibre attach, pigtails, lid seal; reject chipping $>25$\,\textmu m at facet, waveguide cracks, contamination.\\
\rowcolor{rowalt} MIL-STD-883 Method 2010 & Internal (pre-cap) visual inspection & Cavity-package PICs before seal: die-attach, wire/ribbon bond and fibre-attach integrity.\\
MIL-STD-883 Method 1014 & Seal --- fine \& gross leak & Fine leak (He bombing, Cond.\,A) $<5\times10^{-8}$\,atm$\cdot$cm\textsuperscript{3}/s for cavity $<0.05$\,cm\textsuperscript{3}; gross leak by fluorocarbon bubble.\\
\rowcolor{rowalt} MIL-STD-883 Method 2020 & PIND (particle-impact noise detection) & 5 cycles, 20\,g peak, 60--250\,Hz; reject on any noise hit; cavity-package PICs (butterfly, BTF, hermetic LCC).\\
ECSS-Q-ST-60-05C & Generic procurement for hybrids (hermeticity, PIND, screening) & ESA equivalent for packaged InP / SiPh PIC modules.\\
\rowcolor{rowalt} Telcordia GR-468-CORE \S3.3, Tables 4-1/4-2, \S6.7 & Visual, optical baseline \& hermeticity for optoelectronic devices & IL/RL/PDL and spectral baseline; AWG channel spacing $\le5\%$, crosstalk $\le-25$\,dB; beam-combiner $V>0.95$.\\
ECSS-Q-ST-70-08C; IEC 61300 series & Optical-interconnect measurement methods & IL, return loss and PDL measurement at operating $\lambda$ (e.g.\ 600--1700\,nm); RL $>40$\,dB, PDL $\le0.3$\,dB.\\

\multicolumn{3}{@{}l}{\cellcolor{section}\textbf{\textcolor{esablue}{Phase 2 --- Environmental (steps 5--10)}}}\\
GSFC-STD-7000B (GEVS) \S2.4 & Sinusoidal \& random vibration & Random qual 14.1\,g rms, 20--2000\,Hz, 3\,min/axis ($\le22.7$\,kg); sine to 20\,g axial/14\,g lateral; $\Delta$IL $\le0.2$\,dB, no fibre-attach failure.\\
\rowcolor{rowalt} GSFC-STD-7000B \S2.4.5 & Pyroshock & Shock-response spectrum to 4000\,g at 10\,kHz.\\
GSFC-STD-7000B \S2.6.2 & Thermal-vacuum cycling & $\ge4$ cycles (8 for new technology), $\le1\times10^{-5}$\,Torr; in-vacuum optical functional test at each plateau.\\
\rowcolor{rowalt} MIL-STD-883 Method 2002 / 2001 / 2007 & Mechanical shock / constant acceleration / vibration & Shock 1500\,g, 0.5\,ms half-sine (Cond.\,B); constant accel.\ 5000\,g (Cond.\,D)--20\,000\,g (Cond.\,E) on die--package interface.\\
MIL-STD-883 Method 1010 / 1011 & Temperature cycling / thermal shock & $-55/+125$\,\textcelsius\ (Cond.\,C), 500 cyc; PIC-tailored $-40/+85$\,\textcelsius\ (Cond.\,B); $\Delta$IL $\le0.5$\,dB, $\Delta$crosstalk $\le2$\,dB.\\
\rowcolor{rowalt} MIL-STD-883 Method 1004; IEC 60068-2-67 & Moisture resistance / damp heat (THB) & 85\,\textcelsius\ / 85\,\%RH, 1000\,h, biased; non-hermetic SiPh/InP; $\Delta$IL $\le0.5$\,dB, $\Delta$responsivity $\le10\%$.\\
ECSS-E-ST-10-03C \S5/\S5.4/\S5.6; ECSS-Q-ST-70-04C & Mechanical, TVAC \& shock testing; thermal testing & ESA equivalents to the GEVS/MIL mechanical and thermal-vacuum sequences.\\
\rowcolor{rowalt} ECSS-Q-ST-70-02C / ASTM E595 & Thermal-vacuum outgassing screening & 125\,\textcelsius, 24\,h, $10^{-5}$\,Torr; TML $\le1.0\%$, CVCM $\le0.1\%$; critical for fibre-to-chip adhesives.\\

\multicolumn{3}{@{}l}{\cellcolor{section}\textbf{\textcolor{esablue}{Phase 3 --- Reliability \& life (steps 11--12)}}}\\
MIL-STD-883 Method 1005 & Steady-state life (HTOL) & Active PICs 2000\,h at $T_{j,\max}$ ($\sim$85\,\textcelsius), nominal bias; $\ge22$ devices, 0 failures (LTPD 10); $\Delta I_\text{th}\le20\%$, $V_\pi$ drift $\le10\%$.\\
\rowcolor{rowalt} MIL-STD-883 Method 1015 & Burn-in & 168\,h at 125\,\textcelsius\ (Cond.\,A/B); reject $>20\%$ drift in optical output power, threshold or wavelength.\\
Telcordia GR-468-CORE \S6.2 / \S5; JEDEC JESD22-A108 & Endurance \& burn-in / temperature-bias-operating-life for optoelectronics & Arrhenius acceleration $E_a$ 0.4--0.7\,eV (III--V lasers); active-PIC life-test procedure.\\

\multicolumn{3}{@{}l}{\cellcolor{section}\textbf{\textcolor{esablue}{Phase 4 --- Radiation hardness assurance (steps 13--16)}}}\\
ESCC Basic Spec.\ 22900; MIL-STD-883 Method 1019; ASTM F1892 & Total-ionising-dose (Co-60 $\gamma$) steady-state test \& bias/anneal sequence & 30--300\,krad(Si) LEO/GEO, to 1\,Mrad(Si) Jovian; dual dose-rate for ELDRS; track $I_\text{th}$, $I_d$, responsivity, $V_\pi$.\\
\rowcolor{rowalt} ESCC Basic Spec.\ 25100; MIL-STD-750 Method 1017; ASTM E722 & Displacement-damage dose (proton/neutron, NIEL-scaled) & 50--200\,MeV protons to $1\times10^{11}$--$10^{13}$\,p/cm\textsuperscript{2}; InP/Ge dark-current rise $\le2\times$; responsivity loss $\le10\%$.\\
MIL-STD-750 Method 1080; JEDEC JESD57 & Single-event-effects (heavy-ion) test procedure & LET 1--80\,MeV$\cdot$cm\textsuperscript{2}/mg; SEL-immune $\ge37$; pulsed-laser 1064\,nm complementary screening of sensitive nodes.\\
\rowcolor{rowalt} ECSS-Q-ST-70-06C; ASTM E512 & UV / charged-particle exposure of photonic materials & VUV 115--200\,nm + solar UV 200--400\,nm to 2000 ESH; cladding-loss increase $\le0.1$\,dB/cm; no surface darkening.\\
ECSS-Q-ST-60-15C Rev.1 & Radiation-hardness-assurance overarching requirements & Sets RDM $\ge2$ and the combined TID+DDD+SEE plan; depth set by passive/active class (Phase~0).\\

\multicolumn{3}{@{}l}{\cellcolor{section}\textbf{\textcolor{esablue}{Phase 5 --- ESD, EMC \& final acceptance (steps 17--19)}}}\\
MIL-STD-883 Method 3015; JEDEC JS-001 / JS-002 & ESD sensitivity --- HBM / CDM classification & HBM class 1A (250\,V) minimum for III--V active PICs, class 2 (2\,kV) preferred; CDM $\ge250$\,V.\\
\rowcolor{rowalt} GSFC-STD-7000B \S2.5 (tailored MIL-STD-461G); ECSS-E-ST-20-07C & Conducted/radiated emissions \& susceptibility (instrument level) & CE101/CE102, RE101/RE102, CS101/CS114/CS115/CS116, RS101/RS103; magnetic cleanliness for fine-pointing missions.\\
ECSS-Q-ST-10-09C; ECSS-Q-ST-20C; NASA-STD-8739.11 / EEE-INST-002 & Nonconformance, QA \& EEEE parts final acceptance & Repeat baseline vs pre-environmental data: $\Delta$IL $\le1$\,dB (passive), $\Delta\lambda$ $\le50$\,pm; Qualification Test Report + DML/DCL.\\

\multicolumn{3}{@{}l}{\cellcolor{section}\textbf{\textcolor{esablue}{Phase 6 --- Flight demonstration \& heritage (steps 20--22; programmatic, no bench procedure)}}}\\
NPR 8735.1 & Problem reporting / mission-assurance process & On-orbit anomaly capture and corrective action; no coupon test.\\
\rowcolor{rowalt} ESA in-orbit-demonstration / validation (IOD/IOV) practice & On-orbit performance verification & On-orbit optical performance compared to ground baseline; feeds operational heritage back into Detail Specifications.\\

\end{longtable}
\end{landscape}

\subsection{Operationalising the template within ESA and NASA frameworks}
The template is intended to slot into the existing assurance machinery rather
than replace it. On the ESA side, the natural home is a PIC-specific
\emph{annex} to ECSS-Q-ST-60C augmented by a pair of ESCC specifications --- a
Generic Specification covering the seven-phase flow and a family of Detail
Specifications for the individual platforms (silica AWG, Si\textsubscript{3}N\textsubscript{4}
AWG/microcomb, ULI beam combiner, SOI/InP active PIC) --- with qualified
devices entering the ESCC Qualified Parts List via the ESCIES
portal. On the NASA side, the same flow maps onto a PIC
\emph{addendum} to NASA-STD-8739.11 / EEE-INST-002, expressed in the Parts
Assurance Level vocabulary (PAL~4 commercial for Class~D demonstrators
through Class~S screening for flagships). In both cases
the Phase-0 classification decision is what allows a single document to span
the Class~D--to--Class~A range, and the TRL gates of Table~\ref{tab:template}
provide the objective entry criteria that programme reviews (PDR, CDR, QR)
already expect. Because passive astrophotonic cores are radiation-tolerant by
construction, the template's largest practical effect is to redirect scarce
beam time and packaging effort toward the active III--V devices and toward
the five gating gaps identified in \S\ref{sec:platform_outlook}, rather than
toward re-testing inherently robust passive structures.

\subsection{Qualification processes not covered by existing ESA/NASA documents}\label{sec:gaps}
Table~\ref{tab:procedures} maps every PIC-SQT phase onto an \emph{existing}
ESA, NASA or industry test method. A complementary and equally important
question is the converse: which qualification processes are \emph{relevant}
to space PICs but have \emph{no} adequate procedure in the current document
set? Because the ECSS, GSFC-STD-7000B, MIL-STD-883/750, Telcordia and JEDEC
methods were written for discrete EEE/optoelectronic parts, hermetic
packages and optical fibres, several effects that are specific to
\emph{integrated} photonic circuits fall through the gaps. These are
test-method gaps, distinct from the technology-maturity gaps of
\S\ref{sec:platform_outlook}. Table~\ref{tab:gaps} lists the most important,
why the existing documents do not capture them, and the nearest standard or a
proposed verification. They cluster into four themes: (i) optical-domain
metrology that electrical part-screening never measures (coupling-alignment
retention, in-situ radiation-induced attenuation and phase drift, optical
single-event transients); (ii) operating regimes outside the standard test
envelopes (cryogenic operation below the $-65$\,\textcelsius\ floor of
MIL-STD-883 Method 1010, and high on-chip optical-power / laser-induced-damage
handling in vacuum); (iii) space-environment interactions defined for
materials but never applied to PIC waveguides (atomic-oxygen erosion of
exposed cladding, deep-dielectric charging of the SiO\textsubscript{2}/Si\textsubscript{3}N\textsubscript{4}
stack, hydrogen-induced loss); and (iv) synergistic, life-duration effects
(combined UV $+$ thermal-vacuum $+$ radiation ageing of fibre-attach
adhesives, and self-contamination of facets over mission life).

\begin{landscape}
\centering\scriptsize
\setlength{\LTpre}{0pt}\setlength{\LTpost}{0pt}
\begin{longtable}{@{}p{4.6cm} p{8.0cm} p{7.7cm}@{}}
\caption{Qualification processes relevant to space PICs that are \emph{not}
adequately covered by the ESA/NASA documents underlying
Table~\ref{tab:procedures}, with the reason and a proposed verification. These
are test-method gaps that complement the technology-maturity gaps of
\S\ref{sec:platform_outlook}.}\label{tab:gaps}\\
\rowcolor{esablue}\textcolor{white}{\textbf{Missing qualification process / test}} &
\textcolor{white}{\textbf{Why existing ESA/NASA documents do not cover it}} &
\textcolor{white}{\textbf{Nearest standard / proposed verification}}\\
\endfirsthead
\rowcolor{esablue}\textcolor{white}{\textbf{Missing qualification process / test}} &
\textcolor{white}{\textbf{Why existing ESA/NASA documents do not cover it}} &
\textcolor{white}{\textbf{Nearest standard / proposed verification}}\\
\endhead

\multicolumn{3}{@{}l}{\cellcolor{section}\textbf{\textcolor{esablue}{Optical-domain metrology not present in EEE part screening}}}\\
Fibre-to-chip / edge-coupler alignment-retention (sub-\textmu m) & MIL-STD-883 mechanical methods (2002/2007/2001) and the seal/PIND tests judge the \emph{package}; none verifies retention of the sub-\textmu m optical alignment of a fibre-array-unit or edge coupler, the dominant PIC failure mode. & In-situ insertion-loss monitoring \emph{through} vibration, shock and thermal cycling with an explicit $\Delta$-alignment / $\Delta$IL budget (e.g.\ $\le0.2$\,dB); extension of GR-468 mechanical sequences with live optical read-out.\\
\rowcolor{rowalt} Radiation-induced attenuation (RIA) \& refractive-index / phase drift in waveguides & MIL-STD-883 Method 1019 and ESCC 22900 track \emph{electrical} parameters; no method measures in-situ optical loss \emph{and} phase at the operating wavelength versus dose --- yet index drift moves AWG channels and degrades interferometric null depth~\cite{yin2021,piacentini2020}. & In-situ optical RIA and interferometric phase measurement under Co-60 and proton beams; pass criteria on $\Delta$IL, $\Delta\lambda_\text{channel}$ and $\Delta\phi$ at $\lambda_\text{op}$ (cf.\ rad-hard fibre practice, \S\ref{sec:fibers}).\\
Optical single-event transients in active control elements & JEDEC JESD57 / MIL-STD-750 M.1080 capture electrical SEU/SEL/SET; transient phase or intensity glitches in thermo-optic phase shifters, MZI heaters and on-chip lasers (mode-hop, RIN spikes) from ionising strikes are not addressed. & Heavy-ion and pulsed-laser SEE with simultaneous optical-output capture; quantify transient null-depth / channel-power excursions and recovery time.\\

\multicolumn{3}{@{}l}{\cellcolor{section}\textbf{\textcolor{esablue}{Operating regimes outside the standard test envelopes}}}\\
\rowcolor{rowalt} Cryogenic operation \& cycling ($\le80$\,K) & MIL-STD-883 Method 1010 cycling stops at $-65$\,\textcelsius; CTE-mismatch stress, optical performance and repeated deep-cryo cycling of IR PICs are outside all listed methods. & Dedicated 80\,K\,$\leftrightarrow$\,300\,K cycling with in-situ optical functional test (the cryogenic-cycling row of the TRL matrix, \S\ref{sec:trl}); facilities such as CSL FOCAL, NASA GSFC SES.\\
High on-chip optical power / laser-induced-damage threshold (LIDT) in vacuum & EEE standards contain no optical-power-handling or facet-LIDT test; in vacuum there is no convective cooling, so self-heating, two-photon absorption and facet damage at the high intensities used for microcombs and pump light are uncharacterised. & Vacuum LIDT and power-ramp test with thermal imaging; derate on-chip intensity against measured damage and thermal-runaway thresholds.\\

\multicolumn{3}{@{}l}{\cellcolor{section}\textbf{\textcolor{esablue}{Space-environment interactions defined for materials but not applied to PIC waveguides}}}\\
Atomic-oxygen (AO) erosion of exposed cladding / polymer / AR coatings & AO ground testing exists (ASTM~E2089, ISO~15856) but is written for bulk materials; it is not applied to PIC facets, cladding or anti-reflection coatings on ram-facing LEO optics. & AO exposure of representative facet/cladding and coating coupons with before/after throughput and loss metrology at $\lambda_\text{op}$.\\
\rowcolor{rowalt} Deep-dielectric / internal charging of the SiO\textsubscript{2}/Si\textsubscript{3}N\textsubscript{4} stack & ECSS-E-ST-20-06C Rev.1 (spacecraft charging) exists but is not invoked in the PIC flow; thick low-conductivity dielectric stacks can store charge and discharge internally in GEO/HEO/L2 electron environments, a mechanism absent from HBM/CDM ESD tests. & Internal-charging assessment per ECSS-E-ST-20-06C Rev.1 tailored to the multilayer PIC stack; electron-beam charging test with discharge monitoring.\\
Hydrogen-induced loss / index change in silica \& Ge-doped cores & A documented fibre-optic effect (H\textsubscript{2} outgassed by spacecraft materials raises IR absorption) with no counterpart procedure for integrated waveguides. & H\textsubscript{2}-exposure test with loss-at-$\lambda$ monitoring over representative partial pressure and life (per optical-fibre heritage, \S\ref{sec:fibers}).\\

\multicolumn{3}{@{}l}{\cellcolor{section}\textbf{\textcolor{esablue}{Synergistic and life-duration effects}}}\\
Combined-environment (UV $+$ TVAC $+$ radiation) ageing of fibre-attach adhesives & ECSS-Q-ST-70-06C, ASTM~E595 and TID methods test these stresses \emph{singly}; synergistic degradation of UV-cure epoxies and encapsulants at the fibre-chip joint can exceed the sum of single-stress results. & Sequential or simultaneous combined-environment chamber test on representative joints; $\Delta$IL and bond-strength acceptance over mission-equivalent dose/UV/cycles.\\
\rowcolor{rowalt} Long-duration facet throughput vs self-contamination & ASTM~E595 / ECSS-Q-ST-70-02C measure TML/CVCM of materials but not the resulting throughput loss from molecular redeposition on optical facets over mission life in vacuum. & Life test with periodic in-vacuum throughput measurement; contamination budget expressed as $\Delta$throughput at $\lambda_\text{op}$ rather than mass loss.\\

\end{longtable}
\end{landscape}

Several of these gaps are decisive for astrophotonic science cases rather
than merely conservative: radiation-induced index drift and optical
single-event transients act directly on the quantities that interferometric
beam combiners and AWG spectrographs must hold stable (null depth, channel
position), and cryogenic and high-power optical handling gate the IR-spectrograph
and frequency-comb applications surveyed in \S\ref{sec:app_cases}. None
requires a new measurement \emph{principle} --- each can be realised by
adding live optical read-out to an existing environmental, radiation or
contamination sequence, or by tailoring an existing materials standard
(ASTM~E2089, ECSS-E-ST-20-06C Rev.1) to the PIC. They are therefore natural
candidates for the platform-specific Detail Specifications proposed in
\S\ref{sec:results}, and would extend the PIC-SQT of Table~\ref{tab:template}
from a re-use of heritage EEE methods into a genuinely photonic
qualification standard~\cite{jovanovic2023}.

\subsection{Space qualification of photonic wire bonds and two-photon-polymerised structures}\label{sec:pwb}
Photonic wire bonding (PWB) is an emerging hybrid-integration technology in
which a three-dimensional polymer waveguide is written \emph{in situ} by
two-photon polymerization (TPP) directly between the facets of dissimilar
chips --- for example a silicon or Si\textsubscript{3}N\textsubscript{4} PIC,
an InP laser, and a single-mode-fibre array --- adapting the mode fields and
bridging the misalignments that would otherwise demand sub-micron active
alignment~\cite{pwb_lindenmann2012,pwb_blaicher2020}. Because the bond is
printed to the as-measured positions of the components, PWB attacks two of
the gating gaps identified in \S\ref{sec:platform_outlook} and
\S\ref{sec:gaps} --- the fibre-to-chip interface and space-qualified hybrid
III--V integration --- in a single step. The same TPP process also prints
free-form facet couplers, microlenses and mechanical spacers, so the
qualification questions below apply to TPP-printed structures in general, not
only to wire bonds.

As a fabrication class, TPP is very new to spaceflight: to our knowledge
there is no published space-qualification campaign and no ESA or NASA standard
that addresses printed-polymer photonic interconnects. Terrestrial reliability
data are nonetheless encouraging and provide a starting baseline. Hybrid
PIC--laser--fibre engines assembled entirely by PWB have reached telecom-grade
performance and survived accelerated-ageing and thermal-cycling
stress~\cite{pwb_blaicher2020}, and polymer wire bonds have been shown to
couple with $\sim$2\,dB loss per channel and to remain robust down to 5\,K
through repeated cryogenic cycling and cryostat bake-out~\cite{pwb_lin2023} ---
a directly relevant result for the cold IR-instrument regime of
\S\ref{sec:trl}. What is missing is the space-specific evidence.

The dominant qualification concern is that a PWB is a photopolymer, and
polymers are the material class that space-assurance standards scrutinise most
heavily. Mapping TPP structures onto the PIC-SQT of Table~\ref{tab:template}
and the gap analysis of \S\ref{sec:gaps}, the following tests are decisive and
largely uncharacterised for TPP resins:

\begin{itemize}[leftmargin=1.4em,nosep]
  \item \textbf{Outgassing and self-contamination.} The cured resin and any
  unreacted monomer must meet TML $\le1.0\%$ and CVCM $\le0.1\%$ (ASTM~E595 /
  ECSS-Q-ST-70-02C); and because the bond sits \emph{on} the optical path, the
  redeposition of volatiles onto adjacent facets must be bounded as a
  throughput loss --- the life-duration test flagged in \S\ref{sec:gaps}. The
  degree of cure (monomer conversion) governs both outgassing and index
  stability and must be a controlled, verified process parameter rather than a
  by-product of the write.
  \item \textbf{Radiation-induced loss and index change.} Photopolymers can
  cross-link or chain-scission under total-ionising dose and protons, shifting
  refractive index and adding absorption at the operating wavelength; the
  in-situ optical radiation-induced-attenuation and phase test of
  \S\ref{sec:gaps} must be run on representative bonds, since a PWB carries the
  full signal and any radiation-induced loss adds directly to the link budget.
  Unlike the passive silica and Si\textsubscript{3}N\textsubscript{4} cores of
  \S\ref{sec:radsummary}, the polymer bond cannot be assumed radiation-tolerant.
  \item \textbf{UV/VUV and atomic-oxygen durability.} Exposed polymer is
  vulnerable to solar UV/VUV darkening (ECSS-Q-ST-70-06C) and, in LEO ram
  directions, to atomic-oxygen erosion (ASTM~E2089); anything but shielded
  internal placement is likely to require a hermetic or inorganic
  over-cladding.
  \item \textbf{Thermo-mechanical integrity.} Cure shrinkage builds internal
  stress, and the large CTE mismatch between the polymer and the chip or fibre
  it bridges drives $\Delta$IL under vibration, shock and $-55/+125$\,\textcelsius\
  cycling; the coupling-alignment-retention test of \S\ref{sec:gaps} --- live
  optical read-out through the mechanical and thermal sequences --- is the
  natural acceptance criterion.
  \item \textbf{Hermetic encapsulation and moisture.} As a non-hermetic
  organic, a PWB is sensitive to humidity and requires either damp-heat
  screening (85\,\textcelsius/85\,\%RH) or encapsulation within a hermetic
  package at Phase~1 of the PIC-SQT.
\end{itemize}

Because none of these tests requires a new measurement \emph{principle}, TPP
interconnects fit within the proposed framework as a new material class rather
than a new flow: a project would classify the bond as a passive-but-organic
element at Phase~0, add the polymer-specific outgassing, radiation-loss,
UV/atomic-oxygen and cure-control tests above to Phases~1--4, and treat the
alignment retention of the printed bond as a primary end-of-line criterion.
Establishing a documented degree-of-cure specification, together with a first
radiation and outgassing dataset on a standard commercial TPP resin, would
move photonic wire bonding from its present laboratory maturity toward the
TRL-6 flight gate and remove one of the principal packaging obstacles to
flying hybrid-integrated astrophotonic PICs.

\section{Conclusions}\label{sec:conclusions}
Astrophotonic PICs are at the threshold of spaceflight: passive
silica, Si\textsubscript{3}N\textsubscript{4} and ULI platforms already sit
at TRL~5--7 with demonstrated radiation tolerance, the first dedicated
silicon-photonics and ULI space-qualification campaigns are in the open
literature, and four decades of optical-fibre heritage provide the interface
on which PIC packaging can build. What has been missing is not a new test,
but an agreed, tailorable framework that turns the scattered ECSS, GSFC and
EEE-INST requirements into a single defensible flow. The qualification
template of Table~\ref{tab:template} (PIC-SQT), specific steps of Table~\ref{tab:procedures} and Table~\ref{tab:gaps} are offered as that framework.
From it, a concrete roadmap for adapting PICs to space follows:

\begin{enumerate}[leftmargin=1.6em]
  \item \textbf{Adopt the seven-phase template as the baseline flow},
  classifying every device as passive or active and binding it to a mission
  class at Phase~0 (Table~\ref{tab:template}).
  \item \textbf{Concentrate the radiation campaign on active devices.} Treat
  passive silica/Si\textsubscript{3}N\textsubscript{4}/ULI cores with a single
  confirmation test and reserve the full TID/DDD/SEE matrix for III--V and
  Ge-on-Si components~\cite{piacentini2020,mao2024,yin2021}.
  \item \textbf{Drive maturation through the TRL gates} of
  Fig.~\ref{fig:trlcoverage}, using the coverage matrix to decide which
  activities are mandatory at each level.
  \item \textbf{Close the five gating gaps}: hermetic UV-PIC packaging;
  RHA campaigns for ALD alumina, AlF\textsubscript{3} and AlN-on-sapphire;
  cryogenic qualification of Si\textsubscript{3}N\textsubscript{4}/ULI/SOI
  for IR spectrographs; space-qualified hybrid III--V integration; and
  space-grade PDKs~\cite{jovanovic2023}.
  \item \textbf{Build flight heritage through LEO smallsat demonstrators},
  which are simultaneously the cheapest route to TRL~7 and the heritage feed
  for HWO-class flagships (\S\ref{sec:tailoring}).
  \item \textbf{Codify the template within the agencies}: a PIC annex to
  ECSS-Q-ST-60C plus ESCC Generic/ Detail specifications on the ESA side, and
  a PIC addendum to NASA-STD-8739.11 / EEE-INST-002 on the NASA
  side.
  \item \textbf{Iterate the template} as the first flight results return,
  feeding operational heritage (Phase~6) back into the Detail Specifications
  and the radiation-design margins.
\end{enumerate}

\noindent Followed in sequence, these steps convert the present patchwork of
borrowed telecom and EEE standards into a coherent, agency-endorsed
qualification path. The result is a template that ESA and NASA can apply
today to the first generation of space astrophotonic instruments --- from a
CubeSat-hosted AWG spectrograph to the photonic beam combiners and
frequency-comb calibrators envisaged for the Habitable Worlds Observatory ---
and refine as that generation returns its first flight data.

\subsection*{Disclosures}
The author declares that there are no financial interests, commercial
affiliations, or other potential conflicts of interest that could have
influenced the objectivity of this research or the writing of this paper.

\noindent\textbf{AI tool disclosure.}
Portions of the manuscript text were edited with the assistance of Claude (Anthropic) for language, format, grammar, and citation checks. All technical content, scientific judgements,
figures, and data presented in the paper are the sole responsibility of the
author. No AI tool was used to generate or analyse the underlying scientific
data or to produce the figures.

\subsection*{Code, Data, and Materials Availability}
This is a review and qualification-flow paper; no new experimental data
sets or software are released alongside it. The published standards and
specifications cited throughout the paper are accessible through the
respective issuing bodies (ECSS, ESCIES, NASA NEPP and the NASA Technical
Standards portal) at the URLs given in the References section. Any
working calculations underlying the tailoring matrix
(Sec.~\ref{sec:tailoring}) and the TRL-versus-test-coverage roadmap
(Sec.~\ref{sec:trl}) are available from the corresponding author upon
reasonable request.

\subsection*{Acknowledgments}
The authors thank colleagues at the DLR and TU Berlin, for many helpful discussions on space qualifications.

\nocite{*}
\bibliographystyle{spiejour}
\bibliography{references}

\vspace{2ex}\noindent\textbf{Kalaga Madhav}  is the Head of R \& D Astrophotonics(innoFSPEC), at the Leibniz Institute for Astrophysics Potsdam. His group develops photonic technologies for astronomical instrumentation, including integrated spectrographs, OH-suppression filters, photonic lanterns, beam combiners, and frequency combs for precision astronomy.

\vspace{2ex}\noindent\textbf{Aashia Rahman} is a Senior Scientist at the Leibniz Institute for Astrophysics Potsdam (AIP), Germany, with more than 20 years of combined research and industry experience. She received her PhD in Instrumentation and Applied Physics from the Indian Institute of Science (IISc), Bangalore, in 2009. Her research interests include fiber optics in astrophotonics, fiber-to-chip coupling, and the development of specialized filters for ground-based astronomical instrumentation.

\vspace{1ex}\noindent Biographies of the other authors are not
available.

\end{spacing}
\end{document}